\documentclass[sigplan,screen]{acmart}

\copyrightyear{2025}
\acmYear{2025}
\setcopyright{rightsretained} 

\makeatother

\usepackage{xcolor}
\usepackage{annotate-equations}
\usepackage{algorithm}
\usepackage{algorithmic}
\usepackage{graphicx}
\usepackage{tcolorbox}
\usepackage{multirow}
\usepackage{array}

\usepackage{xcolor}
\usepackage{epsfig}
\usepackage{natbib}
\usepackage{tabularx}
\usepackage{algorithm}
\usepackage{mathtools}

\usepackage{wrapfig}
\long\def\comment#1{}
\usepackage{multirow}
\usepackage{wrapfig}
\usepackage{textcomp,booktabs}
\usepackage{colortbl}
\usepackage{multirow}
\usepackage{rotating}
\usepackage{epstopdf}
\usepackage{xspace}
\usepackage{bbm}
\definecolor{Gray}{RGB}{230,230,230}

\usepackage{bbold}
\usepackage{graphicx}
\usepackage{caption}
\usepackage{enumitem}
\usepackage{lipsum}
\usepackage{soul}
\usepackage{color}

\usepackage[normalem]{ulem}
\usepackage[usestackEOL]{stackengine}

\newcommand{\todo}[1]{\textcolor{black}{#1}}

\usepackage[labelfont=bf,textfont=normalfont,textfont=sl]{caption}

\usepackage{balance}
\usepackage{makecell}
\usepackage{sistyle}
\SIthousandsep{,}
\usepackage{filecontents}
\usepackage{tikz}

\usepackage[T1]{fontenc}

\newcommand{\sol}{{DarwinGame}}

\newtcolorbox{highlighted}{colback=yellow,coltext=black,breakable}
\definecolor{red_color}{rgb}{1.0, 0.0, 0.0}
\definecolor{black_color}{rgb}{0.0, 0.0, 0.0}
\definecolor{blue_color}{rgb}{0.0, 0.0, 1.0}

\usepackage{tcolorbox,tikz}
\tcbuselibrary{skins}
\tcbuselibrary{breakable}

\newtcolorbox{mybox}[3][]
{
  breakable, 
  enhanced,
  colback  = #2!5, 
  colframe= #2!5,
  boxsep=-0.5mm,
  borderline west={1.5mm}{0.05mm}{#3!30}, 
  #1,
}

\newtcolorbox{myregbox}[3][]
{
  breakable, 
  enhanced,
  colback  = #2!5, 
  colframe= #2!5,
  boxsep=-0.5mm,
  #1,
}

\newcounter{RQcounter}

\newcounter{IQcounter}

\usepackage{cclicenses}

\begin{document}

\title[\sol{}: Playing Tournaments for Tuning Applications in Noisy Cloud Environments]{\sol{}: Playing Tournaments for Tuning Applications in Noisy Cloud Environments}

\author{Rohan Basu Roy}
\affiliation{
    \institution{University of Utah}\city{Salt Lake City}
  \state{UT}
  \country{USA}
}

\author{Vijay Gadepally}
\affiliation{
    \institution{Massachusetts Institute of Technology}\city{Lexington}
  \state{MA}
  \country{USA}
}

\author{Devesh Tiwari}
\affiliation{
    \institution{Northeastern University}\city{Boston}
  \state{MA}
  \country{USA}
}

\begin{abstract}
\textit{This work introduces a new subarea of performance tuning -- performance tuning in a shared interference-prone computing environment. We demonstrate that existing tuners are significantly suboptimal 
by design because of their inability to account for interference during tuning.  \todo{Our solution, \sol{}, employs a tournament-based design to systematically compare application executions with different tunable parameter configurations, enabling it to identify the relative performance of different tunable parameter configurations in a noisy environment. Compared to existing solutions, \sol{} achieves more than 27\% reduction in execution time, with less than 0.5\% performance variability.} \sol{} is the first performance tuner that will help developers tune their applications in shared, interference-prone, cloud environments. 
}

\end{abstract}



\maketitle

\section{Introduction}
\label{sec:introduction}

\textbf{Background.} Tuning application-level and system-level configurable parameters to minimize application execution time, has been long-studied and is widely accepted to be a challenging problem~\cite{bronevetsky2008clomp,tiwari2011auto,ding2015autotuning,tapus2002active,hollingsworth1999prediction,ansel2014opentuner,hollingsworth2010end,thiagarajan2018bootstrapping,roy2021bliss}. This is because an application performance tuning requires navigating through a vast and often non-linear search space of many application-level and system-level configurable parameters. These tuning parameters interact with each other, making the tuning process challenging and time-consuming~\cite{silvano2016antarex,silvano2016autotuning,sarkar2020ddaring,sarkar2009software,roy2021bliss,liu2021gptune,wu2023ytopt}. 

In response to this challenge, \todo{researchers and practitioners} have innovated many performance tuners\todo{~\cite{balaprakash2018autotuning,
yu2020hyper,wu2019paraopt,ansel2012siblingrivalry,
li2024fasttuning,chen2015angel,ansel2014opentuner,yang2023cotuner}}. However, all prior tuners implicitly assume that the tuning process is performed in a dedicated computing environment without performance interference from co-located applications. Unfortunately, users are increasingly running their applications in shared environments including cloud computing platforms due to a variety of reasons (briefly outlined below), and the inability to account for interference during tuning makes existing tuners suboptimal (reasons briefly outlined below and quantified in Sec.~\ref{sec:evaluation}).

\vspace{3mm}

\noindent\textbf{Increasing needs and challenges of performance tuning in shared, interference-prone execution environments.} The cloud computing platforms have experienced dramatic adoption in the last decade~\cite{varghese2018next,wittig2023amazon,destefano2023cloud}. Even when organizations use on-premise clusters, possibly as a supplement to cloud computing, resources are often shared among users to save operational costs, promote more efficient usage of resources, and reduce queue wait time~\cite{simakov2016quantitative,ucbnodesharing,utahhpc,copik2024software,sdsccondo,xiong2018tangram}. 


There are various other reasons why application developers may increasingly use production cloud platforms to tune their applications' performance. For example, certain computing resources (e.g., a specific chip or AI accelerator) may only be available on cloud computing platforms~\cite{reuther2020survey}. Production cloud provides a diverse range of resources and computing environments, that are available on-demand, avoiding the need for significant upfront investments in dedicated on-premise clusters~\cite{buyya2018manifesto,
delimitrou2016hcloud}. The end user may not be able to afford dedicated on-premise resources and rather use pay-as-you-go renting of shared resources (on the cloud computing platform)~\cite{alam2020cloud,shafiei2022serverless}. Additionally, the application may be intended to be deployed on a cloud computing ecosystem and hence, should be developed and performance-tuned on the same platform type to ensure a particular configuration's sensitivity to interference is captured during tuning~\cite{gupta2011evaluation,klein2014brownout}. \textit{Therefore, we need tuners that can be effective when applications are executed on interference-prone environments.}


One intuitive solution is to use existing tuners in the shared cloud environment and rely on its outcome. While simple, our experiments reveal that existing tuners are significantly suboptimal (more than 40\% far from optimal) when used in shared, interference-prone execution environments (Sec.~\ref{sec:evaluation}). This is expected because existing tuners are not aware of and are designed to account for the impact of interference in a shared execution environment  -- \textit{e.g.}, an effective tuning configuration may be at a serious disadvantage if the interference is particularly high during its evaluation. 


\textit{Unfortunately, designing tuners in a shared cloud environment is particularly challenging because the effect of performance interference from co-located applications cannot be actively controlled or manipulated in a production cloud-based execution environment}. Furthermore, the level of performance inference or noise on such platforms varies significantly, making it difficult to accurately assess the true effectiveness of a particular tuning configuration. One naive solution is to rent dedicated computing infrastructure on the cloud just for tuning purposes -- but, they are often prohibitively expensive~\cite{zhang2020high}, the tuning process may last weeks/months~\cite{balaprakash2018autotuning}, and determining a configuration without interference-awareness may lead to tuning configuration that is sensitive to noise and exhibit high-performance variation in a shared cloud setting (as shown in Sec.~\ref{sec:background}). \textit{Therefore, the goal of this work is to design and implement a tuner that works effectively and efficiently in shared, interference-prone environments.}

\vspace{3mm}

\noindent\textbf{The following are the contributions of \sol{}.} 

\vspace{1.5mm}
\noindent\textbf{Introduction of \todo{a new research area and first-solution}.} Currently, application developers do not have access to any capability that can perform performance tuning in an interference-prone cloud-based environment. \textit{\sol{} is the first solution to address this challenge.}

\vspace{1.5mm}
\noindent\textbf{Novel tournament-based tuning in a shared execution environment.} \sol{} bypasses the inability to manipulate or control the performance interference on shared cloud computing hardware, by co-locating multiple copies of the same application with different parameter tuning configurations to \textit{approximately} assess the relative potential of different parameter configurations with respect to each other, and their sensitivity to background interference (referred to as ``playing games''). However, as Sec.~\ref{sec:design} demonstrates, these co-located ``games'' need to be organized and played systematically to pick \todo{a winner}. \sol{} employs a multi-phase tournament strategy where each phase is played in multiple rounds and with different styles (e.g., Swiss, double elimination, and barrage) to progress toward finding the most promising tuning configuration ( Sec.~\ref{sec:design}).


\vspace{1.5mm}

\noindent\textbf{\todo{Evaluation}.} Our extensive evaluation (including on the AWS platform, using Redis, GROMACS, FFmpeg, and LAMMPS~\cite{carlson2013redis,abraham2015gromacs,tomar2006converting,thompson2022lammps}) demonstrates \sol{}'s effectiveness and its complementary nature to existing tuners. \sol{} reduces the execution time of applications by more than 27\% compared to other existing solutions -- while limiting the performance variation to less than 0.5\%. It can be integrated with existing tuners to improve their effectiveness by more than 15\%. \sol{} is available at: \href{https://doi.org/10.5281/zenodo.14014918}{https://doi.org/10.5281/zenodo.14014918}.

 \section{Background and Motivation}
\label{sec:background}

\noindent\textbf{Terms and Definitions.} In the context of application tuning, \textit{tunable parameters} are application- and systems-level parameters that can be configured to take on multiple values or states. These configurable parameters form the application's \textit{tuning search space} (or search space for short), whose dimension is equal to the number of tunable parameters. The size of the search space is the number of possible combinations of configurations of the tunable parameters. For example, if the search space is $n-$dimensional, and the $i^{th}$ parameter can have $a_{i}$ number of possible values, then the size of the tuning search space is $\prod_{i=1}^{n} a_i$. 

A \textit{tuning configuration} refers to one point in the tuning search space (a configuration has one specific value for each possible parameter). The execution time of an application depends on the chosen configuration. Executing and measuring the execution time of an application with a given configuration is called \textit{sampling}. 

Application tuners sample different tuning configurations to find the configuration with the least execution time. The optimal configuration in a dedicated computing environment, simply referred to as \textit{optimal configuration} (or optimal), is the best tuning configuration, \textit{i.e.}, the one with \textit{the minimum execution time in an interference-free environment (e.g., an on-premise cluster with dedicated nodes)}. Accurately determining the optimal configuration requires sampling all the points of the search space, one by one, in a dedicated environment (not in shared hardware or a performance interference-prone cloud-based execution environment). Determining the optimal configuration is often not feasible in practice, but we performed extensive experiments in dedicated-non-shared environments to determine the optimal configuration for understanding the effectiveness of \sol{}. 


We define the strategy of sampling all the points in the search space in \textit{a shared, cloud-based VM (noisy environment)}, one by one, as the \textit{exhaustive search}. \textit{Note that the exhaustive search process may not yield the same answer as the optimal configuration because the exhaustive search is performed in an interference-prone cloud-based execution environment.} We use an exhaustive search strategy as a comparison point because if a programmer has access to only a cloud-based execution environment, this is the only feasible brute-force method to find an effective tuning configuration. 

The goal of an effective tuner is to determine the optimal configuration (a tuning configuration with the least execution time in the non-shared environment) without performing an exhaustive search in the cloud environment -- to save both time and money. In addition to this, the application tuner should pick a tuning configuration that experiences low-performance variations across different runs (less performance sensitivity to noise or interference). In other words, in a cloud-based environment, \sol{} needs to find a tuning configuration that results in minimum execution time (possibly, as low as the optimal configuration in a non-shared environment) and low sensitivity to interference. 

\vspace{1.5mm}

\begin{figure}
    \centering
    \includegraphics[scale=0.33]{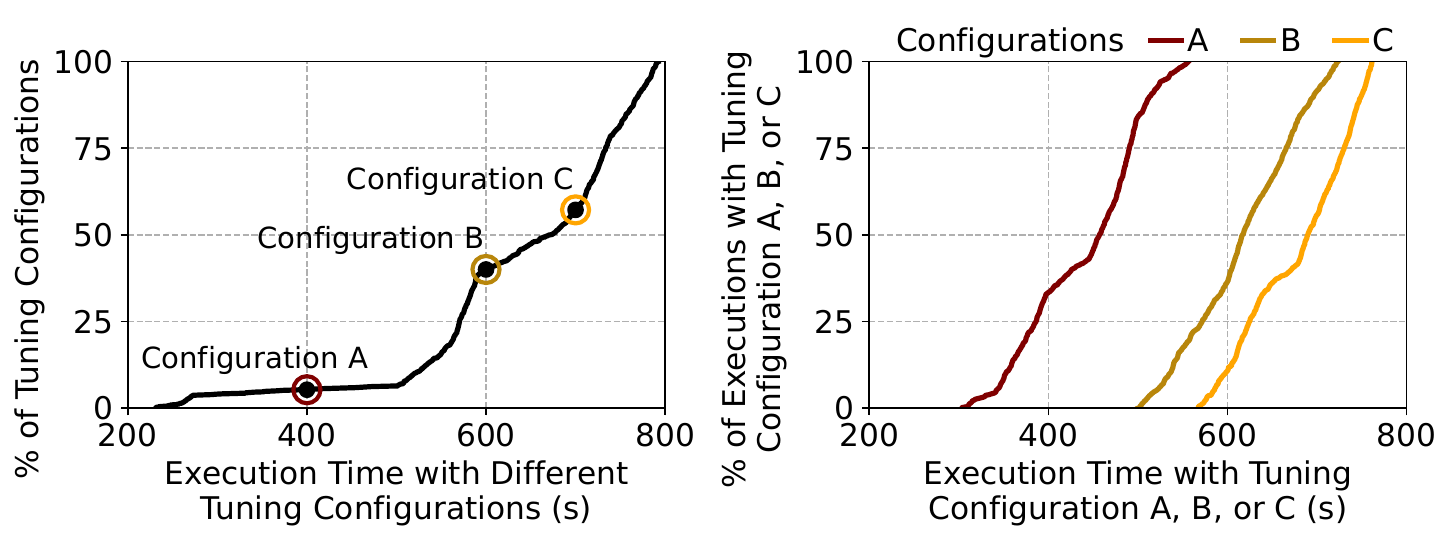}
    \vspace{-3mm}
    \caption{Different tuning configurations result in significantly different execution times (Redis serving one million requests on AWS general purpose m5 VM). Due to the interference-prone execution environment in the cloud, application execution with any given tuning configuration exhibits variation in execution time across different runs.}
    \label{fig:diff_params_motiv}
    \vspace{-2mm}
\end{figure}

\noindent\textbf{Motivation.} As expected, the size of the tuning search space can be large. For example, the size of the search space of Redis with application- and systems-level parameters is 7.8 million (details of parameters and execution in Sec.~\ref{sec:methodology}).  Fig.~\ref{fig:diff_params_motiv} (left) shows the cumulative distribution of execution times for serving one million requests on Redis using 250 randomly chosen tuning configurations. Different tuning configurations result in significantly different execution times -- the range of execution time varies from 230 seconds to 792 seconds. More than 93\% of the tuning configurations result in twice the execution time compared to an execution with the tuning configuration resulting in the least execution time.


The problem becomes even more challenging because the execution time of an application with a particular chosen tuning configuration varies in a cloud-based execution environment, due to external interference and noise in the shared environment. From Fig.~\ref{fig:diff_params_motiv} (right), we observe that when an application is executed 1000 times in the cloud using each of the three chosen tuning configurations (A, B, and C, with average execution time 440s, 617s, and 678s, respectively), their execution time can vary even by 45\%, due to differences in the interference level and varying resource contention from other co-located applications. In other words, the true potential of a tuning configuration cannot be determined without running a prohibitively large number of times.

\begin{myregbox}{yellow}{}
\textbf{Takeaway I.} \textit{As expected, different tuning configurations for any given application yield very different execution times (more than 3$\times$ differences). Unfortunately, due to noise and interference, determining the optimal tuning configuration (the least execution time) in a cloud-based environment is challenging because even the same tuning configuration may exhibit more than 40\% variations across multiple runs.} 
\end{myregbox}

\begin{figure}
    \centering
    \includegraphics[scale=0.32]{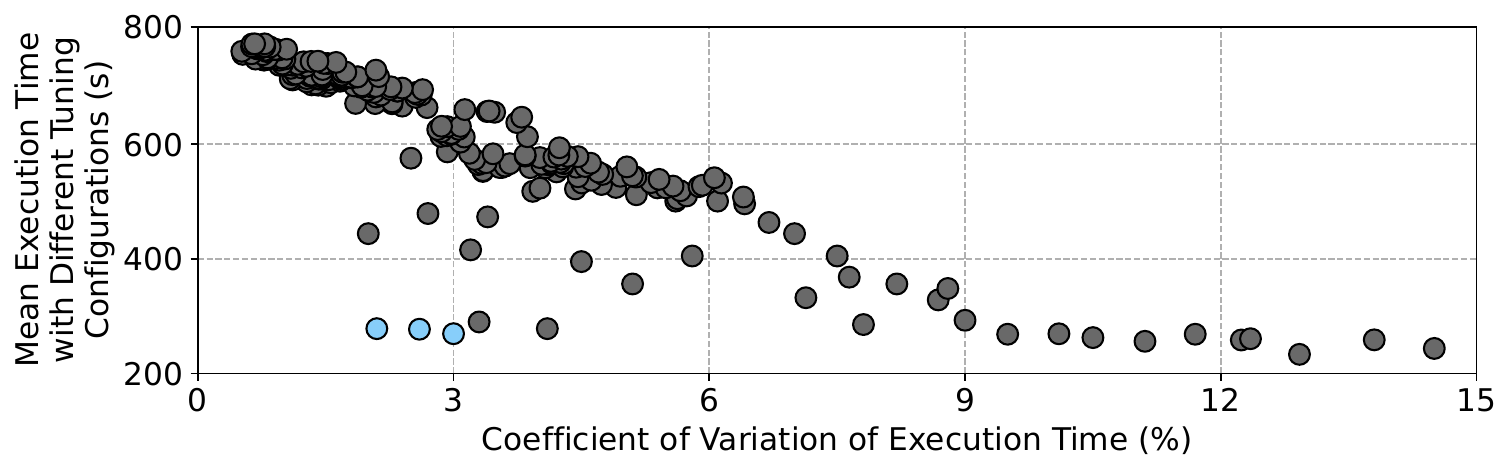}
    \vspace{-3mm}
    \caption{Different tuning configurations exhibit different degrees of performance variation across runs. Achieving only low execution time via tuning is not sufficient. Both low-performance variation and low-execution time are desired -- for example, configurations denoted by blue markers (Redis serving one million requests on AWS general purpose m5 VM). \todo{}}
    \label{fig:cov_motiv}
    \vspace{-2mm}
\end{figure}

To further elaborate on this point, in Fig.~\ref{fig:cov_motiv}, we show a scatter plot of the coefficient of variation of the execution time of 250 randomly selected tuning configurations (1000 runs with each configuration in the cloud) on the x-axis and the mean execution time on the y-axis. We note that tuning configurations yield significantly different performance in the cloud due to interference. Also, the performance variation affects the configurations differently, \textit{i.e.,} the coefficient of variation is different for the different tuning configurations. 

Particularly, we note that there is a trend of application execution using better-performing tuning configurations (lower execution time), to have a higher coefficient of variation than application execution using tuning configurations resulting in higher execution time. This is because, when an application is highly optimized, it is likely that the execution is pushing the system closer to its resource limits or is more sensitive to specific resources. Hence, even small fluctuations in resource availability can cause significant variations in performance. Since after tuning, these applications will be executed in a shared cloud environment, the tuner should find tuning configurations that result in relatively stable performance (less coefficient of variation in execution time). 

However, there are still some tuning configurations (shown in blue in Fig.~\ref{fig:cov_motiv}), which have a low coefficient of variation in execution time and also significantly low mean execution time. These tuning configurations are good candidates for selection by the application tuner. They present an opportunity for a high-performant (low execution time) and stable (low variation in performance) application execution. Since such tuning configurations are less in number, it is challenging for a tuner to find them. We observe similar behavior in terms of mean execution time and performance variation of different tuning configurations for all the evaluated applications (details in Sec.~\ref{sec:evaluation}).  

\begin{myregbox}{yellow}{}
\textbf{Takeaway II.} \textit{The interference-prone and noisy cloud execution environment affects tuning configurations differently -- application executions with certain tuning configurations experience more performance variations than others. A desirable performance tuner should find tuning configurations with low execution time and low-performance variations under interference.}
\end{myregbox}

\begin{figure}
    \centering
    \includegraphics[scale=0.3]{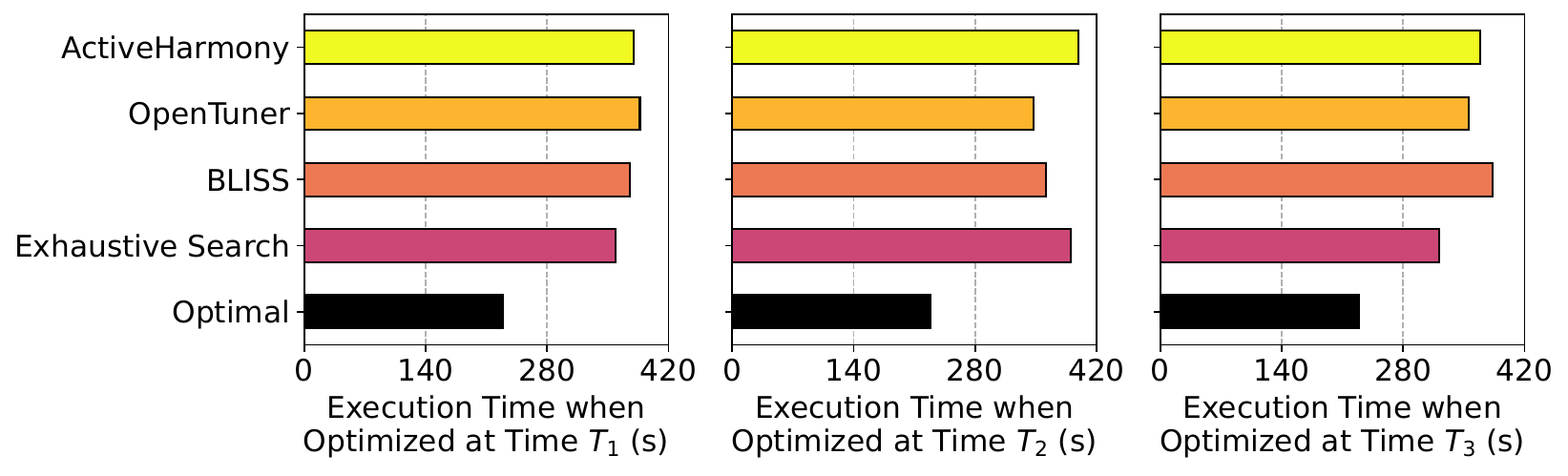}
    \vspace{-3mm}
    \caption{Due to interference posed by the noisy tenants in the cloud, state-of-the-art tuners perform sub-optimally and cannot determine the configuration with the least execution time. Further, their selected tuning configuration is inconsistent -- it changes over time, depending upon the time of the tuning and corresponding interference (Redis serving one million requests on AWS general purpose m5 VM).}
    \label{fig:tuner_motiv}
    \vspace{-2mm}
\end{figure}

State-of-the-art performance tuners focus on tuning applications and executing them in a dedicated environment to reduce execution time. They do not consider performance variation in the cloud during tuning and execution, and interference from noisy neighbors.  Therefore, naively using existing state-of-the-art performance tuners in a cloud-based environment yields tuning configurations (as their outcome) that are far from the optimal configuration obtained in a non-shared environment. 

Fig.~\ref{fig:tuner_motiv} shows that the execution time of the optimized configuration obtained by performing tuning in the cloud using existing tuners as-is (OpenTuner~\cite{ansel2014opentuner}, ActiveHarmony~\cite{hollingsworth2010end}, and BLISS~\cite{roy2021bliss}).  \todo{These solutions perform tuning and application execution in a cloud-based noisy environment.}  The execution times are much higher than the optimal configuration obtained in a dedicated, interference-free environment. We noted that, while not shown in  Fig.~\ref{fig:tuner_motiv},  performing tuning in a dedicated, interference-free environment using existing tuners as-is, indeed results in execution times very close to the optimal configuration obtained in a dedicated, interference-free environment -- highlighting that existing tuners become ineffective purely because of uncontrollable and unregulated interference in the \todo{cloud-based execution environments. The interference in the cloud changes between the executions of an application with different tuning configurations, which leads the existing tuners to misinterpret the relative performance of different configurations.} 

One may suspect that exhaustively searching all tuning configurations in the cloud may be more effective (Exhaustive Search). While reasonable, even exhaustive search results in a significantly higher execution time than the optimal configuration's execution time. As expected, this is because of the effect of varying degrees of interference on different tuning configurations. 

Furthermore, Fig.~\ref{fig:tuner_motiv} also demonstrates that when tuning is performed multiple times in the cloud during different time intervals, the application tuners select different configurations and as a result, the relative effectiveness of the tuners varies among multiple rounds of tuning. This happens because the application execution time with a tuning configuration changes with a change in the interference level in the cloud at different time intervals when tuning is performed. 

\begin{myregbox}{yellow}{}
\textbf{Takeaway III.} Existing state-of-the-art performance tuners, when used as-is, are ineffective in a cloud-based environment because their design is fundamentally unaware of the interference experienced in a cloud-based environment and its effect on various tuning configurations -- resulting in producing significantly suboptimal tuning configurations.
\end{myregbox}

Motivated by these takeaways, next, we propose the design of \sol{}.

\section{Design of \sol{}}
\label{sec:design}
In this section, we first set the goal of \sol{} and discuss the overview of the design.

\subsection{Design Goal}
\label{des_goal}
\sol{}'s design goal is to find tuning configurations that result in \textit{(a) low execution time and (b) low variations in performance during execution in a cloud-based environment, and (c) the low tuning time} as tuning time contributes toward application development time and cost. 


In production-shared cloud environments, the level of interference from other tenants can not be deterministically controlled or configured by the end user and hence, its impact on the evaluation of a particular tuning configuration cannot be assessed. To bypass this, \textit{\sol{} employs the co-location of multiple copies of the same application with different tuning configurations, on the same node(s) to determine their relative potential.} In the context of \sol{}, this is referred to as playing games (co-located execution of an application with different tuning configurations) between different players (different tuning configurations). 


The key insight is that instead of attempting to accurately measure the true execution time (when no interference is present) of each tuning configuration, \sol{} attempts to get approximate relative ranking and possibly, improve this approximation by trying multiple times/games among promising tuning configurations. 


\sol{} plays a tournament of multiple games in multiple phases to judge the relative performance of good-performing configurations multiple times, with multiple opponent tuning configurations, before deciding on the most desirable tuning configuration. Playing a tournament of multiple games helps \sol{} to accurately judge the relative performance of different tuning configurations, even when the interference level varies.  However, the tournament structure needs to be carefully designed to ensure that \sol{} achieves its tuning goals and finds configurations with the desirable properties (design goals (a), (b), \& (c)).

\begin{figure}
    \centering
    \includegraphics[scale=0.72]{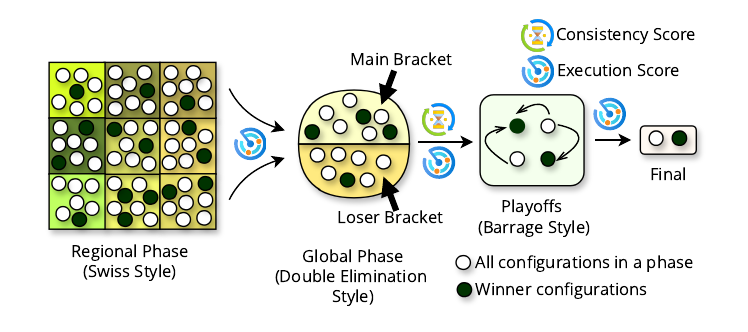}
    \vspace{-3mm}
    \caption{Overview of \sol{}'s tournament-based tuning: multiple phases and different playing styles.}
    \label{fig:overview}
    \vspace{-2mm}
\end{figure}

\subsection{Design Overview}
\label{des_overview}

\todo{\textbf{Game formats of \sol{}'s tournament.} \sol{}'s tournament-based design is based on three major formats of playing games: Swiss, double elimination, and barrage. In the Swiss format~\cite{hua2017swiss}, players compete across several rounds and are paired against others with similar performance scores, allowing for a ranking-based progression. The double elimination format~\cite{dinh2020simulating} offers participants a second chance by moving them to a loser bracket if they lose once; they face elimination only after a second loss, ensuring fair opportunities for recovery. The barrage style~\cite{key} is used toward the final stages, where the best-performing players from previous rounds compete in a series of direct knockouts. This format also allows participants a brief opportunity to recover if they lose early and ensures that only the strongest tuning configurations progress to the final round.}

\vspace{3mm}

\noindent\textbf{Tournament structure.} As visually depicted in Fig.~\ref{fig:overview}, \sol{} structures the tournament to select application tuning configuration in four phases (each phase can have multiple rounds and each phase has a different playing style -- that is, how configurations compete against each other and how they advance). 

In the first phase, \sol{} divides the entire parameter search space into different regions and plays the regional phase in a Swiss tournament style~\cite{hua2017swiss}. In this tournament style, the most promising players directly compete with each other and are given chances to progress further. This tournament style is chosen for this phase because the Swiss tournament style is most suited for identifying players (application execution with best-performing tuning configurations) with the highest potential among a large pool of players (the first phase has the largest number of players). Sec.~\ref{des_regional} describes more design trade-offs and how \sol{} navigates to ensure that if some regions have more potential, then more winners are selected from them. The winners from each region then proceed to the global phase. 

In the global phase, the tournament is played in a double elimination style~\cite{dinh2020simulating} -- so that the losing tuning configurations get an additional opportunity to remain in the tournament without being directly eliminated -- because these configurations have proven their worth in the past and \sol{} attempts to avoid sudden elimination due to ``one bad day (high interference sensitivity)''. 

In the global phase, \textit{tuning configurations are judged based on their execution time and variations in performance}. This is done to determine tuning configurations that are consistent in performance and do not exhibit significant performance sensitivity to the noise/interference in a cloud-based execution environment. The winners of the global phase, along with the winner among all the losing configurations of the global phase, proceed to the \textit{playoffs phase} which is played in the barrage style~\cite{key}. The top two tuning configurations from the playoffs play in the final to determine the winner.

\todo{In each game of the tournament, \sol{} co-locates multiple copies of the same application with different tuning configurations and runs them in parallel to identify their relative performance and variability in the presence of noise. Empirically, we observed this approach outperforms other strategies where each configuration is individually exposed to the background noise or controlled interference tests (often by more than 10\%). This improvement arises from \sol{}'s ability to expose all competing tuning configurations to identical noise conditions, allowing for fair and consistent comparison. This form of comparison is required due to the unpredictable nature of cloud environments and the inability to control the background noise.}

\todo{When performing performance tuning in shared environments, fluctuating background interference makes it difficult to compare application execution with different parameter configurations accurately over time. Methods like Bayesian optimization, which rely on proximity-based variance estimates, become less effective when nearby samples face different noise levels. Similarly, statistical methods like quantile regression and Thompson sampling, which are often used to handle variability, are also unable to account for unpredictable cloud interference (resulting in significantly less effective results compared to \sol{}). By simultaneously evaluating multiple configurations under the same conditions, \sol{}'s tournament structure effectively adapts to interference, ensuring that comparisons are fair and reflective of real-world cloud environments. This approach reduces bias from noise induced fluctuations and systematically narrows down the best-performing configurations, making it more robust than traditional optimization techniques in noisy, shared environments.}


\todo{We also highlight that unpredictable background interference also makes it challenging to provide any provable theoretical bounds for \sol{}. This is why \sol{} is, on purpose, designed to be driven by intuitive choices, instead of claiming any theoretical bounds about its efficacy. We remark that the theoretical underpinning of tournament design is widely regarded as a difficult problem in multiple communities~\cite{ryvkin2008predictive,rajkumar2021theory,
harbring2003experimental,fuhrlich2022improving} – especially in terms of providing any proven theoretical bounds because of the uncertainty and variance-related challenges. As expected, prior studies largely focus on analytical trade-offs between fairness, complexity, incentive, etc., and often ultimately rely on probability-driven computer simulations under various simplistic assumptions~\cite{csato2021tournament} instead of hard provable bounds.}

\vspace{3mm}

\begin{figure}
    \centering
    \includegraphics[scale=0.58]{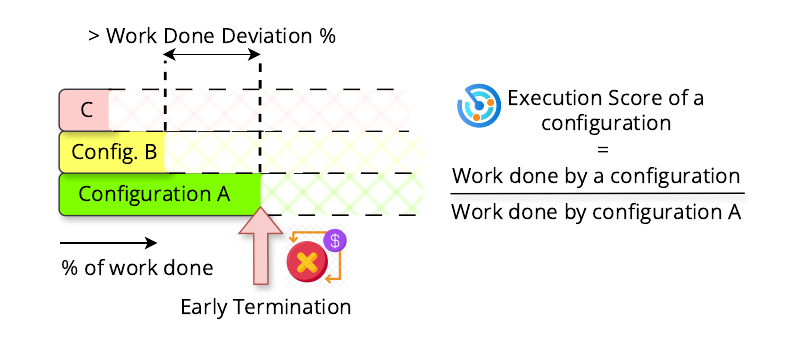}
    \vspace{-3mm}
    \caption{\sol{} performs early termination in the initial phases of the tournament to reduce tuning time (Configuration A has the highest execution score).}
    \label{fig:exe_score}
    \vspace{-2mm}
\end{figure}

\noindent\textbf{\sol{}'s strategies for reducing tuning time.} Recall from Sec.~\ref{des_goal} that one of the design goals of \sol{} is to reduce the tuning time. To achieve this, \sol{} makes several adjustments in the structure of the games played in the tournament -- (1) in the initial rounds of the tournament, multiple players (more than just two) can play together. (2) The games allow \textit{early termination}, which means that when a player is performing significantly well compared to the other players, the game gets terminated and the winner is declared. \sol{} keeps track of the amount of work being completed by each of the players while they are competing together in a game. 

Now, when \sol{} observes that the difference in the work done between the fastest and the next best player is more than the work done deviation percentage ($d$), \sol{} terminates the game and declares the winner even before the players have finished execution (Fig.~\ref{fig:exe_score}). (Sec.~\ref{sec:methodology} discusses the methodology to determine the amount of work done). We set the value of $d$ to be 10\%. However, we observe less than 2.7\% difference in \sol{}'s execution-time-wise outcome when $d$ is varied between 5\% to 15\%. \sol{} allows early termination when at least 25\% of the total work done is completed by the fastest configuration. (3) Also, \sol{} makes arrangements to play multiple games in parallel within a phase. Next, we will discuss the details of the regional phase.

\begin{figure}
    \centering
    \includegraphics[scale=0.30]{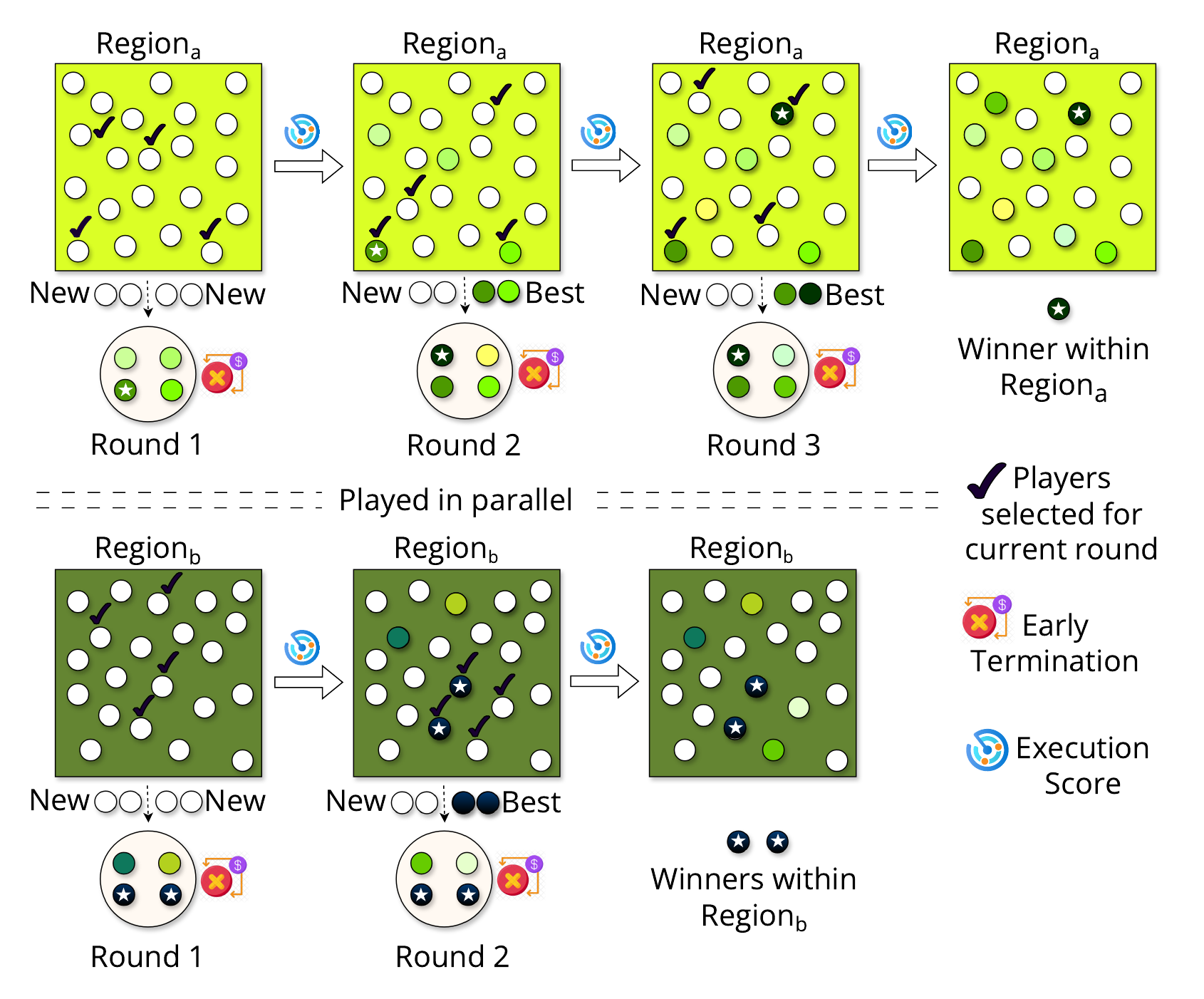}
    \vspace{-3mm}
    \caption{The regional phase is played in a Swiss style to quickly identify the tuning configuration(s) with the highest potential. White circles denote configurations that are not yet selected to play within any round. A darker shade of a configuration (which has played in a particular round) denotes better performance. A white star inside a circle denotes the winner configuration(s) of a round.}
    \label{fig:regional}
    \vspace{-2mm}
\end{figure}

\subsection{Phase I : Regional Phase of \sol{}}
\label{des_regional}
The entire search space size can be in the range of millions. To tackle such a large number of players (tuning configurations), \sol{} divides the entire search into $n_r$ number of different regions. In our experiments, we set $n_r$ to be 10,000. However, \sol{}'s execution time-wise result varies by less than 3.7\%, when $n_r$ is varied between 5000 to 15,000 across all evaluated applications. All the points from the $n$-dimensional search space of an application are mapped to a one-dimensional index. Based on the index values, the tuning configurations corresponding to the different points are assigned to different regions, so that all regions contain an equal number of tuning configurations. 

Inside each region, the number of players that play a game together is simply chosen equal to be number of cores ($P$), but is configurable. Making multiple players compete together helps to reduce the tuning time compared to playing more games with only two players. Note that, \sol{} cannot play a game between all the players to decide the winner as the number of tuning configurations is in the range of millions and all of them can not be co-located in a node. Even when we play games multiple times between the maximum number of most promising tuning configurations that can be co-located (1000 configurations), the resulting winner is far from the optimal solution (more than 2.8$\times$ more execution time on average). This is because co-location inside a VM creates additional noise, impacting the performance of promising tuning configurations. 


Games in different regions can be played in parallel in different VMs to reduce tuning time. Within each region, the games are played for multiple rounds in a Swiss style. In a Swiss-style tournament, the increasingly most promising players compete with each other. This form of tournament helps to eliminate low-potential players quickly and identify the most promising ones. Since the goal of \sol{} in this initial phase of the tournament is to quickly identify the most potent tuning configurations in a large search space, it adopts a Swiss playing style. Also, all the games in this round allow for early termination to save tuning time. The relative performance of each tuning configuration in a game is determined via the \textit{execution score}. It is the fraction of work done in a game by an application execution with a tuning configuration compared to execution with the fastest tuning configuration, until the game is completed (the fastest execution completes) or is terminated early (Fig.~\ref{fig:exe_score}). 

\begin{figure}
    \centering
    \includegraphics[scale=0.62]{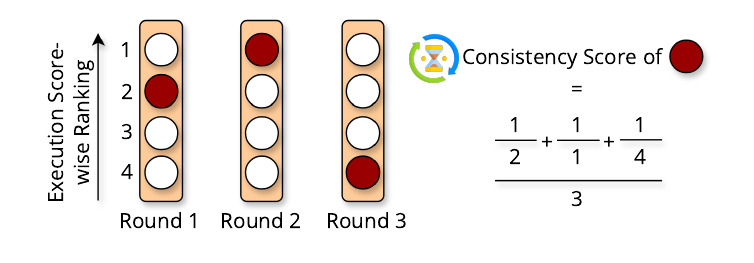}
    \vspace{-3mm}
    \caption{A high consistency score denotes low-performance variations of tuning configurations with high execution scores.}
    \label{fig:consistency_score}
    \vspace{-2mm}
\end{figure}

In the first round within a region, players are selected randomly to play a game, and thereafter, assigned an execution score. \textit{From the next rounds, half of the players are selected from the player pool that has never been selected to play a game (new players; no assigned execution score), and the other half is chosen from players with assigned execution scores (best players from the ones that have played a game before).} This selection is performed probabilistically -- a higher execution score means more probability of being selected to play a game. Thus, players with high scores contend with each other, making the tournament progress in Swiss style (easier to figure out the best-performing players among a large group of players).  

Within each region, if \sol{} observes that one player is consistently winning for more than one time, it terminates the regional stage and declares the winner. It proceeds the players within the deviation $d$\% (work done deviation percentage) of the best-performing player in terms of the execution score to the global phase. This is done to make sure that promising players getting negatively impacted by contention from others are still given a chance to proceed in the tournament. If the termination condition of the regional phase is not met, \sol{} continue the rounds of games in each region till each player receives an equal probability of being selected to play at least one game. This termination criterion speeds up the tuning time. The number of rounds played depends on how competitive the players in a region are. When some players are significantly more competitive than others in a region, the winners are determined in a lesser number of rounds.

Fig.~\ref{fig:regional} describes the progress in the regional phase in two different regions. The rounds inside the regions progress with the selection of half of the players based on their execution score (shaded circles), and the other half from the players that have not played a game yet (white circles). As the fastest configuration (circle with a white star) consistently wins rounds more than once, \sol{} terminates the regional phase and declares the winner.  Also, if there are other players, consistently performing within the work done deviation percentage of the fastest configuration, then they are also declared as the winner, and they proceed to the global phase.  If a region has more promising candidates (on average, some regions can be more promising than others), \sol{}'s design ensures it can have multiple winners that advance to the next round. Next, we discuss the global phase of \sol{}.

\begin{figure}
    \centering
    \includegraphics[scale=0.62]{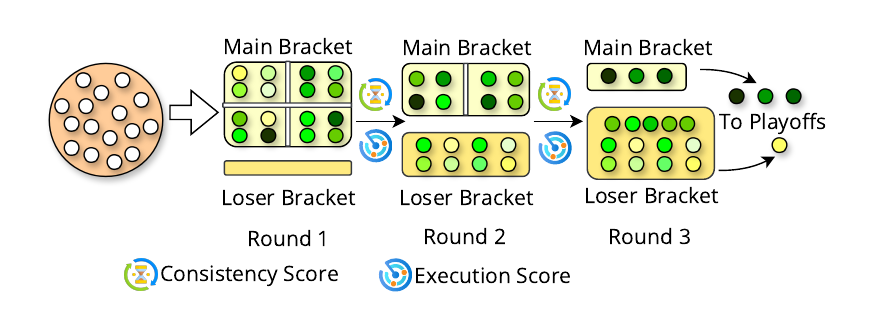}
    \vspace{-3mm}
    \caption{The global phase is played in a double elimination style to prevent losing tuning configurations from being immediately eliminated.}
    \label{fig:global}
    \vspace{-2mm}
\end{figure}

\subsection{Phase II : Global Phase of \sol{}}
\label{des_global}
This phase of the tournament is played in a double elimination style. The phase has two different game brackets -- the main bracket and the loser bracket. In this tournament format, the losers of games are not eliminated immediately. They are transferred to the loser bracket and get an additional chance to proceed in the game. When multiple application executions with different tuning configurations are co-located, certain promising tuning configurations can be negatively impacted to a greater extent by the presence of other executions in the same execution environment. Thus, the double elimination format gives them an additional opportunity to win and proceed in the tournament. The tuning configurations that proceed to the global phase from the regional phase are already promising choices in terms of their execution time. Hence, in this round, they are additionally judged based on the consistency of the performance in the presence of other application executions and background noise in a cloud-based platform. 

When a fixed number of players ($P$) play a game together, we define the \textit{consistency score} of a player as the average of 1/ranking of the player in terms of the execution score from the current and also all the previous games (Fig.~\ref{fig:consistency_score}). This captures the consistency of performance of a tuning configuration over the different games it has played -- a higher consistency score means lower variations in performance of configurations with high execution scores, in a noisy cloud-based environment. After the termination of each game, players are ranked based on their execution score and consistency score. Thereafter, the players are ranked based on the summation of execution score ranking and consistency score ranking. The one with the lowest summation value is declared as the winner.

As shown in Fig.~\ref{fig:global}, The global phase also consists of multiple rounds of games, where winners from the previous round proceed to the next round. Within each round, players form groups of $P$ number of players, and the games among different groups within a round can be played in parallel. \textit{When forming these groups of players within a round, \sol{} ensures that the players within each group initially belonged to different regions of the search space to improve diversity within each playing group.} The winners from each group, within each round, proceed to the next round and stay in the main bracket of games, while all the other players (losers) are transferred to the loser bracket. As the number of rounds progresses, the main bracket size starts decreasing while the loser bracket size starts increasing. The number of rounds continues till the main bracket has only a chosen number of players (we set this number to be three; \sol{}'s chosen tuning configuration that wins the final remains the same even if we increase this chosen number of players beyond three). These players proceed to the next round, the playoffs. All games within the global phase also follow early termination to reduce the tuning time.

Thereafter, among all the players of the loser bracket, $P$ number of players are selected based on their average execution score and consistency score, seen so far. The winner of this game among the players of the loser bracket, receives a \textit{wild card entry} to proceed to the playoffs. Playing additional games in the loser bracket does not result in improvement of \sol{}'s tuning, but increases the tuning time. The tournament structure ensures that all the tuning configurations that proceed to the playoffs have low execution time and variations in performance in the cloud. 

\subsection{Phases III \& IV: Playoffs and Final Phase}
\label{des_playoff_final}
From the global phase, only promising players, in terms of both execution time and performance consistency, proceed to playoffs. To ensure high accuracy in determining the winner tuning configurations, in the playoffs and in the final, the games are played among two players only at a time (the number of co-location is two), and \sol{} does not perform early termination (the games are played till the completion of an application execution with the faster among the two competing tuning configurations). This is because the near-winner configurations are not likely to have significant differences in their execution time. 

The playoff is played in a barrage style, which is a common way to perform the penultimate round of the tournament, as it is a robust form of playing style to ensure only the top contenders proceed to the finals~\cite{camberwellpetanqueclubMatchFormats}. Among the players of the playoffs, the top half of the players with the highest average execution score play with each other. The winners of these games proceed to the next round, while the losers play with the winners of the games between the bottom half of the players in the playoffs, in terms of execution score. For example, if there are four players in the playoffs, the top two players with the highest average execution score play the first game of the playoffs. The winner, in terms of execution score, directly goes to the final. Since all the players in the playoffs have low variations in performance, they are judged based on their execution time. The second game is played between the remaining two players of the playoffs. The loser of this game gets eliminated. Thereafter, the winner of the second game and the loser of the first game plays the third game of the playoffs. The winner of the third game becomes the second finalist of \sol{}.  

The two finalists play the final game of \sol{}. The application execution with the tuning configuration that finishes earlier among the two competing tuning configurations, is the winning tuning configuration. The tournament structure of \sol{} is designed to find tuning configurations that lead to low execution time and low variations in performance.  \sol{} can also be integrated with existing tuners to improve their effectiveness. 

\begin{figure}
    \centering
    \includegraphics[scale=0.58]{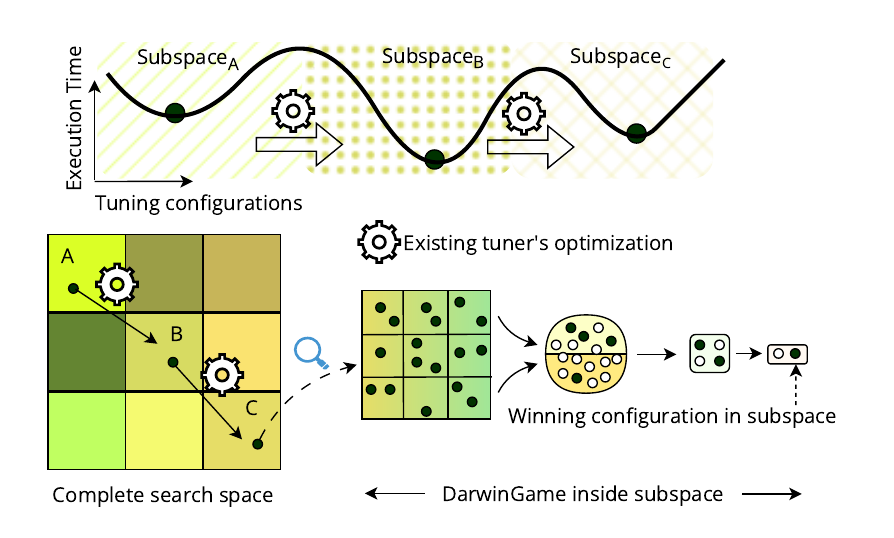}
    \vspace{-3mm}
    \caption{\sol{} can be integrated with existing application tuners by playing tournaments inside subspaces.}
    \label{fig:integration}
    \vspace{-2mm}
\end{figure}

\todo{As noted earlier, \sol{} is designed to be driven by intuitive choices around playing tournaments; therefore, we purposely avoid claiming theoretical bounds. Nevertheless, \sol{}'s design provides flexibility to finish certain rounds within a reasonable number of trials, ensuring efficient convergence. For example, the Swiss-style tournament is expected to converge logarithmically in terms of the the number of rounds -- though the exact rounds required can vary based on factors like relative player rankings, sensitivity to background noise, and player performance variability. Similarly, the global phase's convergence depends on the distribution of consistency scores, which, in turn, reflects the competing players' resilience to interference across rounds. While we do not necessarily force \sol{} to finish in a specified number of rounds, it can be adjusted to achieve specific properties, if desired. Finally, phases III and IV (playoffs and final phase) are configured to converge in a linear number of trials since configurations compete in a straightforward knock-out fashion.}

\vspace{1mm}

\subsection{Enhancing existing tuners with \sol{}}
\label{des_robust}
\sol{} is designed in a way such that it is amenable to be integrated with other existing tuning solutions to improve their effectiveness of tuning in the cloud, reduce the tuning time, and determine tuning configurations that have low variations in performance in the cloud.

\todo{As visually depicted in Fig.~\ref{fig:integration}, this integration is performed by dividing the entire search space into different subspaces. The existing tuner observes each subspace as a single tuning configuration, using its optimization logic to traverse from one subspace to another. Inside every subspace, \sol{} plays a full tournament by dividing the subspace into multiple regions, proceeding through the regional phase, followed by the global phase, playoffs, and the final.}

\todo{This modular approach allows \sol{} to operate within each subspace independently, identifying optimal configurations without disrupting the existing tuner's broader search process. The performance of the winning configuration within a subspace is then used to represent the performance of that subspace, helping the existing tuner make more informed decisions. Sec.~\ref{sec:evaluation} demonstrates that the integration of \sol{} with existing tuners enhances the tuning effectiveness and reduces both tuning time and the resources required to perform tuning.}

\todo{Integrating \sol{} with existing tuners enhances optimization without requiring changes to the pipelines of systems and applications already using those tuners. Its modular design makes integration relatively easier and reduces tuning time in cloud environments. This results in more consistent performance assessments, making \sol{} an effective enhancement to existing frameworks.}

\vspace{3mm}

\noindent\textbf{Putting it all together:} As summarized in Algorithm 1, \sol{} employs a multi-phase tournament structure to optimize application tuning in noisy cloud environments. It begins with regional Swiss-style tournaments to quickly identify top configurations, progresses to a double elimination global phase to ensure consistency under interference, and concludes with barrage-style playoffs and a final to accurately determine the best tuning configuration. \todo{Co-locating application executions with different tuning configurations across multiple games allow \sol{} to accurately assess and rank configurations under noisy conditions. \sol{}'s evaluation across games of the tournament preserves the mathematical basis of fitness measures while improving its robustness in tuning by revealing consistency across progressively challenging tournament phases.} Next, we discuss \sol{}'s evaluation methodology.


\begin{algorithm}[t!]
\caption{\sol{} Tournament for Tuning}
\begin{algorithmic}
\STATE \textbf{Input:} Search space $S$, Regions $n_r$, Players $P$
\STATE \textbf{Output:} Optimal configuration $C^*$

\STATE Divide $S$ into regions $R_1, R_2, \ldots, R_{n_r}$

\FOR{\textbf{each} $R_i$ \textbf{in parallel}}
\STATE Play Swiss style in $R_i$
\STATE $W_i \gets \text{Winner}(R_i)$
\ENDFOR

\STATE Global Phase $G \gets {W_1, W_2, \ldots, W_{n_r}}$

\WHILE{rounds remain}
\FOR{\textbf{each} group $G_j \subset G$ \textbf{in parallel}}
\STATE Play game in $G_j$, calculate scores
\STATE $G_{j,winner} \gets \text{Winner}(G_j)$
\STATE Transfer losers to loser bracket
\ENDFOR
\ENDWHILE

\STATE $P_{main\_bracket\_winners} \gets \text{Winners from } G$
\STATE $P_{wildcard} \gets \text{Winner from loser bracket}$
\STATE $P_{playoffs} \gets P_{main\_bracket\_winners} \cup P_{wildcard}$

\STATE $P_{finalists}$ $\gets$ winners of $P_{playoffs}$ (barrage style)
\STATE Final: $C^* \gets \arg\min_{C \in P_{finalists}} \text{execution\_time}(C)$

\end{algorithmic}
\end{algorithm}

\section{Evaluation Methodology}
\label{sec:methodology}

\begin{table*}[t]
\centering
\begin{tabular}{| p{1.75cm} | p{7.43cm} | p{5.4cm} | p{1.75cm} |}
\hline
\textbf{\small{Application}} & \textbf{\small{Application-level Parameters}} & \textbf{\small{Systems-level Parameters}} & \textbf{\small{Search Space Size}} \\
\hline
\small{Redis~\cite{carlson2013redis}} & \small{tcp-backlog, rdbcompression, rdbchecksum, maxmemory, maxmemory-policy, appendonly, appendfsync, no-appendfsync-on-rewrite, auto-aof-rewrite-percentage, auto-aof-rewrite-min-size, lazyfree-lazy-eviction, lazyfree-lazy-expire, lazyfree-lazy-server-del, hz, dynamic-hz, active-defrag, active-defrag-threshold-upper, \&  active-defrag-cycle-max}
& \multirow{4}{5.5cm}{\small{processor affinity, I/O scheduler, read-ahead setting, vm.swappiness, vm.dirty\_ratio, vm.overcommit\_memory, vm.overcommit\_ratio, vm.dirty\_background\_ratio, vm.dirty\_expire\_centisecs, kernel.sched\_migration\_cost\_ns, kernel.timer\_migration, kernel.sched\_autogroup\_enabled, kernel.sched\_min\_granularity\_ns, kernel.sched\_wakeup\_granularity\_ns, kernel.sched\_rr\_timeslice\_ms, kernel.sched\_rt\_period\_us, kernel.sched\_rt\_runtime\_us, \& kernel.sched\_latency\_ns}} & \small{7.8 million} \\
\cline{1-2} \cline{4-4}
\small{GROMACS~\cite{abraham2015gromacs}}& \small{integrator, nstlist, ns\_type, fourier\_spacing, cutoff-scheme, \& coulombtype} &  & \small{3.8 million} \\
\cline{1-2} \cline{4-4}
\small{FFmpeg~\cite{tomar2006converting}}& \small{compiler optimization levels, function inlining, vectorization, vectorization cost, prefetching, loop unrolling, link-time optimization, stack realignment, -ffast-math, -fomit-frame-pointer, -fstrict-aliasing, -floop-block, -floop-interchange, \& -floop-strip-mine} &  & \small{6.1 million} \\
\cline{1-2} \cline{4-4}
\small{LAMMPS~\cite{thompson2022lammps}}& \small{neighbor skin distance, neighbor list build frequency, timestep, output frequency, integrator, \& cutoff distance} &  & \small{4.4 million} \\
\hline
\end{tabular}

\caption{List of application-, systems-level parameters, and the size of search space of evaluated applications.}
\vspace{-6mm}
\label{table:param_table}
\end{table*}

\noindent\textbf{Applications.} We evaluate \sol{} with four applications: Redis, GROMACS, FFmpeg, and LAMMPS -- used widely for demonstrating the effectiveness of auto-tuners~\cite{gratl2019autopas,bao2019actgan,akiba2019optuna,hampton2010optimal}.

\todo{Redis 6.0} (serving one million requests with Redis benchmark)~\cite{carlson2013redis} is a high-performance, NoSQL database management system, capable of handling various data structures like strings, hashes, lists, and sets with high speed and reliability. GROMACS (with water-cut benchmark)~\cite{abraham2015gromacs} is a versatile molecular dynamics simulation software renowned for its efficiency, scalability, and advanced simulation capabilities. GROMACS is extensively used in computational biophysics, chemistry, and related fields. FFmpeg (with H.264 video file of 10 GB)~\cite{tomar2006converting} is a widely used multimedia framework for handling audio, video, and other multimedia files, offering multiple capabilities for format conversion, streaming, and editing. We tune the compilation parameters and flags for FFmpeg. LAMMPS~\cite{thompson2022lammps} is a molecular dynamics simulation software, recognized for its robustness in handling diverse computational tasks across various scientific disciplines. \todo{All the evaluated applications are multi-threaded; \sol{} does not impose restrictions on the thread counts of applications to be optimized. Instead, it allows each application to use its optimal threading configuration, thereby adapting to the application’s inherent parallelism.}


Table~\ref{table:param_table} lists the specific application-level parameters that we tune and the size of the search space, combining both application-level and systems-level parameters. Different values of these parameters directly impact the execution time and performance variations in the cloud, without affecting the operation or the scientific experiment being performed. 

\todo{Note that, \sol{} aims to optimize for only static tunable parameters. For example, in multimedia processing applications like FFmpeg, these parameters include compiler optimization levels, loop unrolling, and vectorization flags. These parameters are set once during compilation and directly impact the application performance across various hardware configurations. \sol{} focuses on refining these static settings to achieve minimal execution time in interference-prone cloud environments, without runtime adjustments. \sol{} does not directly attempt to dynamically adjust parameters in the middle of execution. However, it can extended to have such capabilities by changing the tournament structure to introduce feedback loops in the global, playoffs, and final phases.}

\vspace{3mm}

\noindent\textbf{Experimental testbed.} We perform the main set of experiments with \text{m5.8xlarge} general-purpose VMs on AWS. We set the value of $P$ (number of players playing a game together in the regional and global phases) to 32 (number of vCPUs in m5.8xlarge). We also evaluate \sol{}'s effectiveness on other AWS VMs of different sizes like m5.large (2 vCPUs), m5.2xlarge (8 vCPUs), m5.16xlarge (64 vCPUs), and m5.24xlarge (96 vCPUs). Additionally, to demonstrate \sol{}'s effectiveness in VMs of different resource configurations, we evaluated \sol{} with other classes of VMs on AWS like c5.9xlarge (compute-optimized instance), r5.8xlarge (memory-optimized instance), and i3.8xlarge (storage-optimized instance). Table~\ref{table:param_table} lists the systems-level parameters we tune via \sol{}. We use \text{taskset} and \text{sysctl} to change the different scheduling and kernel settings affecting the systems-level parameters.

Recall from Sec.~\ref{des_overview}, to perform early termination of games, \sol{} needs to keep track of the percentage of work completed by an application execution with different tuning configurations.  This is achieved by keeping track of the progress in processing or the amount of output produced.  For example, while evaluating Redis, we keep track of the percentage of requests being completed. While evaluating FFmpeg, we keep track of the percentage of frames processed. In GROMACS and LAMMPS, we keep track of the percentage of output produced with respect to the total output size. 

\vspace{3mm}

\noindent\textbf{Competing and complementary solutions.} We compare \sol{}'s performance with other state-of-the-art application tuners: (a) \textit{ActiveHarmony}~\cite{hollingsworth2010end} is a widely used open-source tuning framework that uses a controller and load balancer to navigate through the search space of an application to determine the effect of changing parameter values on the application performance. (b) \textit{OpenTuner}~\cite{ansel2014opentuner} is a state-of-the-art tuning solution that chooses a tuner type based on the geometry of the search space, from a set of pre-existing application tuners (effective and simple). (c) \textit{BLISS}~\cite{roy2021bliss} is a recent tuner that uses an ensemble of Bayesian optimization kernels models to navigate through complex search spaces (effective, but variance in noisy environments).

We show how \sol{} can be integrated with existing solutions to improve their effectiveness -- by dividing the search space into subspaces and running \sol{}’s tournament within each subspace to identify optimal configurations under realistic noise conditions. Also, we compare \sol{}'s performance with (d) \textit{optimal} (optimal configuration determined in a dedicated environment),  and (e) \textit{exhaustive search} (exhaustively searching all configurations in a cloud-based environment), as described in Sec.~\ref{sec:background}. These solutions are practically infeasible but they provide a measure to determine the performance upper bound.

\vspace{3mm}

\noindent\textbf{Evaluation metrics.} We measure the execution time of an application in a cloud-based VM with a tuning configuration to determine execution time-wise tuning performance. Additionally, we repeat an application execution 100 times with the tuned configuration and report the coefficient of variation of execution time as a measure of the variations in performance of a tuning configuration. Since, tuning within different phases can be parallelized, \sol{} uses the core-hours required to perform tuning as a measure of the amount of resources and time used to tune an application.

\section{Evaluation and Analysis}
\label{sec:evaluation}

\begin{figure}
    \centering
    \includegraphics[scale=0.315]{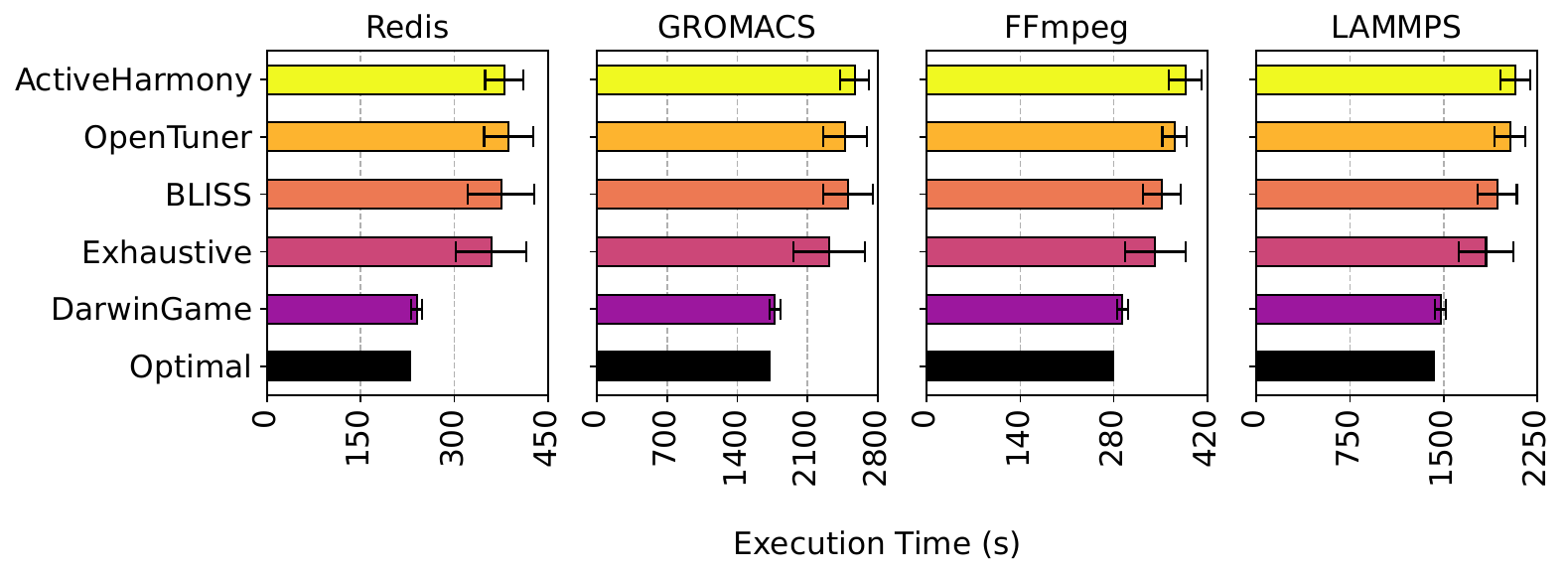}
    \vspace{-8mm}
    \caption{\sol{} is closer to Oracle than other competing solutions. The error bars denote the range of execution time variation when tuning is repeated 100 times separately.}
    \label{fig:tuner_main}
    \vspace{-2mm}
\end{figure}

\noindent\textbf{Can \sol{} perform more effective auto-tuning in the interference-prone cloud-based execution environment, especially compared to existing interference-unaware performance tuners?} Recall from Sec.~\ref{sec:background} (Fig.~\ref{fig:tuner_motiv}) that in the interference-prone execution environment, existing solutions perform poorly compared to the optimal solution because they are not designed to incorporate the effect of noise and resource contention during the performance tuning process. 

\sol{}'s design helps it to bridge the gap between optimal and existing solutions -- in particular, two design features: playing games together to measure the relative performance of different tuning configurations and the tournament structure which ensures that the top contenders are tested multiple times to assess their true potential. From Fig.~\ref{fig:tuner_main}, we observe that on average, \sol{} results in only 4.2\% more execution time than optimal, while the next best tuner (BLISS), results in more than 40\% increased execution time than the optimal solution. 

Since the tuning configurations playing a game together are expected to be impacted by a similar amount of noise in the cloud, \sol{}'s relative measurement of application execution time with different tuning configurations is more consistent than other solutions. This is why, as seen in Fig.~\ref{fig:tuner_main}, the variation range in the execution time (error bars), when \sol{}'s tuning is repeated 100 times, is significantly less than other solutions. In fact, out of the 100 times, \sol{} picks the same tuning configuration 93 times as its final output, while the next best tuner picks 42 different tuning configurations. 

\todo{Note that, it is possible for \sol{} to tune dynamically adjustable parameters like thread count of applications by tweaking the tournament structure to introduce feedback loops in later phases (e.g., global phase or playoffs), where the system dynamically re-ranks configurations based on their performance after adjustments during application execution. However, \sol{} does not perform this dynamic parameter tuning because, in our experimental evaluation, this approach often significantly increased the time and resources used for tuning (over 10\%) for limited performance improvements (less than 5\%).} 

\vspace{3mm}

\begin{figure}
    \centering
    \includegraphics[scale=0.315]{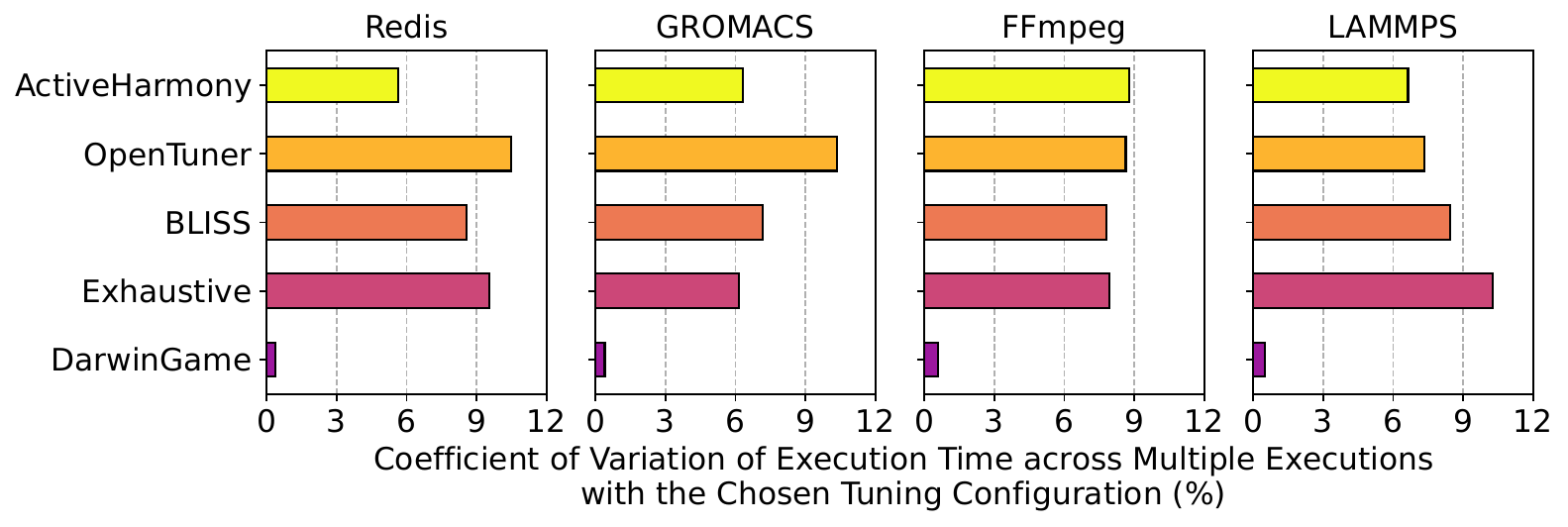}
    \vspace{-8mm}
    \caption{Application execution with \sol{}'s tuned configuration results in lower variations.}
    \label{fig:tuner_cov}
    \vspace{-2mm}
\end{figure}

\noindent\textbf{Does \sol{}'s picked tuning configuration exhibit a lesser degree of performance variation compared to alternative solutions?}  \sol{} picks tuning configurations that exhibit less variation in performance across multiple runs in the cloud.
\todo{The purpose of \sol{} is to execute tuned applications in the cloud without the need to actively control the type of co-located applications. So, while cloud interference distribution shifts are possible, several design components of \sol{} aim to make it resilient to such varying levels of interference.}  

As shown in Fig.~\ref{fig:tuner_cov}, \sol{} leads to lesser variations in performance. The coefficient of variation in execution time in the cloud with \sol{}'s chosen tuning configurations is 0.46\%, while for all other solutions, it is more than 6\%, when an application is executed 100 times with the chosen tuning configuration, at different periods of time, in the cloud. This is because, throughout the tournament, \sol{} makes several efforts to ensure that the chosen tuning configuration has fewer variations in performance -- that is, less sensitivity to performance interference in the cloud. These efforts include playing multiple games with promising configurations across various phases with different opponents. This ensures that the performance of the chosen tuning configuration remains consistent across multiple executions with different levels of contention and noise. Also, \sol{} takes into account the consistency score while deciding the winners of the global phase.

\vspace{3mm}

\begin{figure}
    \centering
    \includegraphics[scale=0.315]{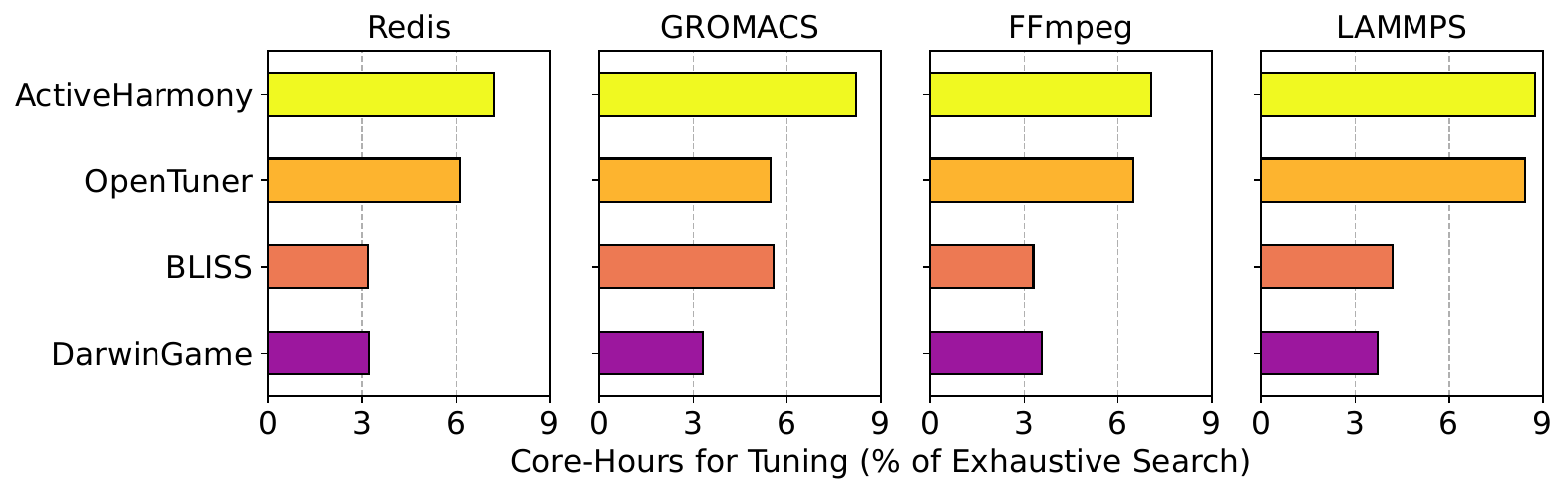}
    \vspace{-4mm}
    \caption{\sol{}'s resource requirements for tuning are less than other state-of-the-art solutions, in most cases.}
    \label{fig:tuner_corehours}
    \vspace{-2mm}
\end{figure}

\noindent\textbf{How does the amount of resources consumed during \sol{}'s tuning compare to existing solutions?} Fig.~\ref{fig:tuner_corehours} shows the core-hours required by \sol{} to complete the tuning process. Since the exhaustive search requires the most resources to perform the tuning, the required core-hours of all other solutions are expressed as a percentage of the exhaustive search.  Exhaustive search requires 9124, 27124, 12778, and 25523 core-hours on m5.8xlarge VM for Redis, GROMACS, FFmpeg, and LAMMPS, respectively. 

We observe that in most cases, \sol{}'s tuning resource requirements are lesser than other solutions. This is because of \sol{}'s resource-aware design components such as early termination of games and playing games with more than two players in the initial phases of the tournament.  In some cases, the required core-hours of \sol{} are slightly higher than a competing solution because of more runs required to assess the interference-sensitivity of certain tuning configurations. \todo{The resource-aware design of \sol{} also helps it to reduce tuning costs of applications in the cloud; the tuning cost can be amortized over 142, 107, 121, and 136 runs of tuned Redis, GROMACS, FFmpeg, and LAMMPS, respectively.}

\vspace{3mm}

\begin{figure}
    \centering
    \includegraphics[scale=0.31]{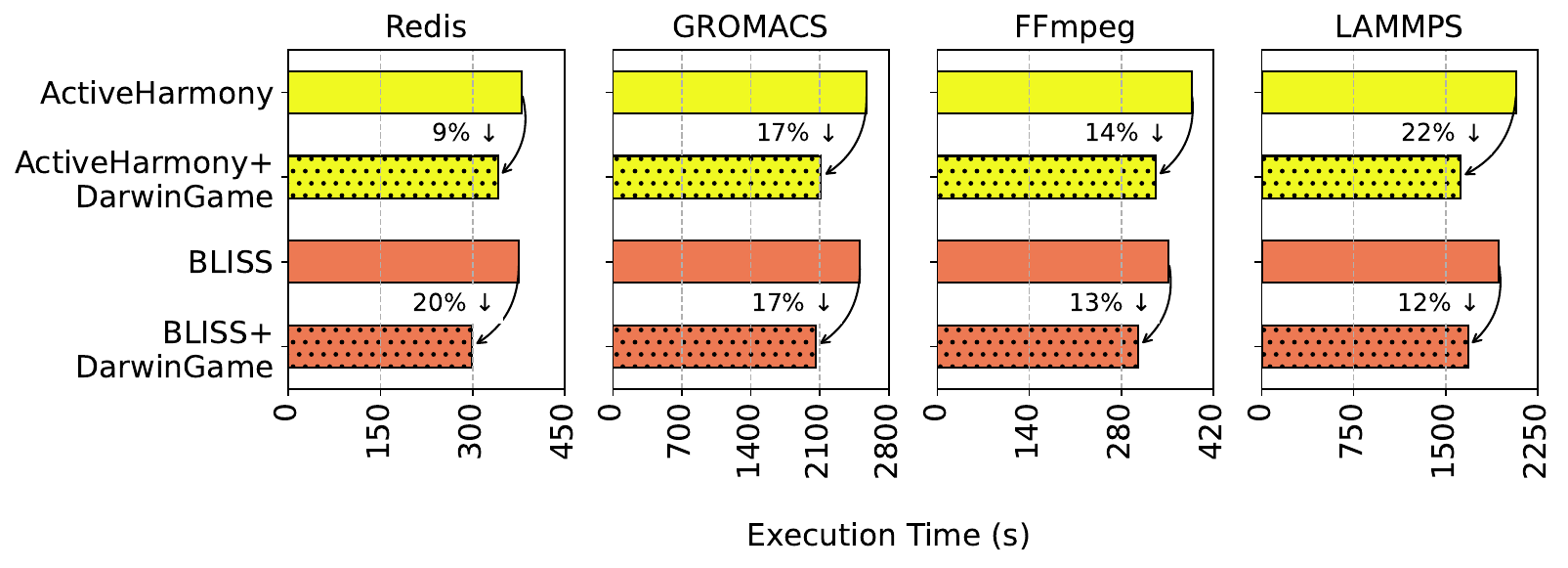}
    \vspace{-8mm}
    \caption{\sol{} can be integrated with existing solutions to improve their effectiveness.}
    \label{fig:tuner_exe_others}
    \vspace{-2mm}
\end{figure}

\begin{figure}
    \centering
    \includegraphics[scale=0.31]{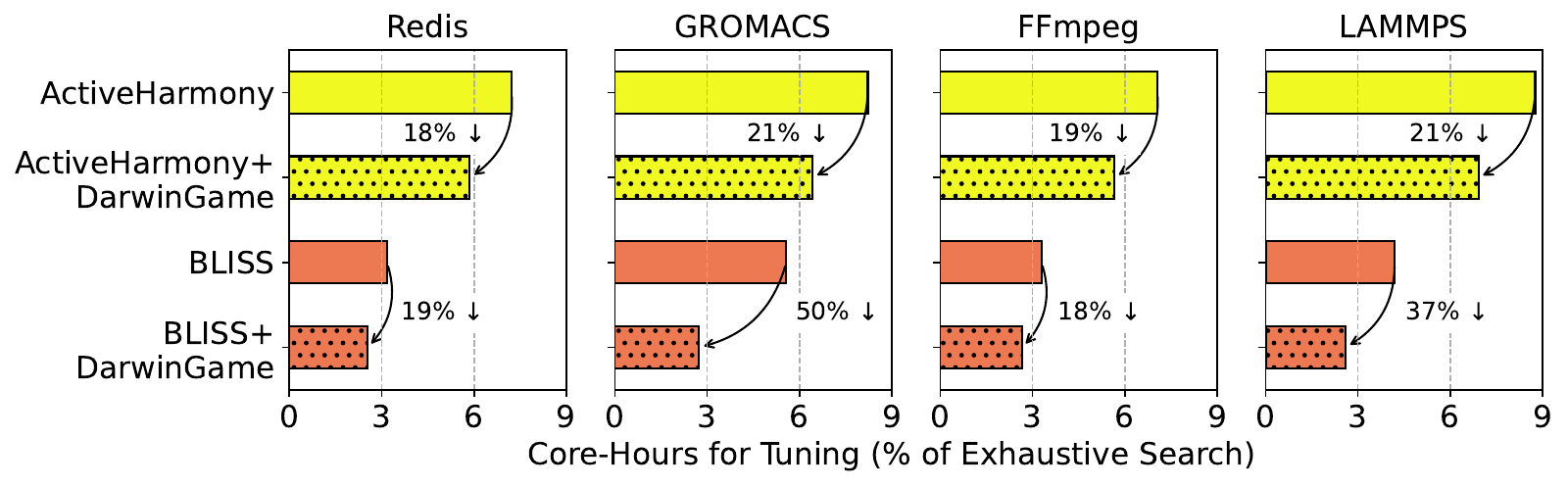}
    \vspace{-8mm}
    \caption{\sol{} can be integrated with existing solutions to reduce their tuning resource requirements.}
    \label{fig:tuner_corehours_others}
    \vspace{-2mm}
\end{figure}

\noindent\textbf{Can \sol{} improve the effectiveness of existing auto-tuners in a complementary manner?}  In Sec.~\ref{des_robust}, we discussed how \sol{}'s tournament-based tuning strategy can be integrated with existing tuning solutions to improve their effectiveness (except OpenTuner due to the non-compatible nature of the search process). 

From Fig.~\ref{fig:tuner_exe_others}, we observe that \sol{} can be integrated with ActiveHarmony and BLISS to improve the application execution time-wise performance by more than 15\%, on average. This improvement is observed because playing \sol{}'s tournament inside each region selected via an existing tuner's optimization logic helps to better estimate the relative performance of application execution in the cloud with different tuning configurations, as the noise in the cloud changes. Due to \sol{}'s early termination and reducing the number of sampling by playing games with multiple players, benefits are also observed in terms of the resource requirements, when \sol{} is integrated with the existing solutions (Fig.~\ref{fig:tuner_corehours_others}). 

\sol{} takes into account the consistency of performance while judging winners and plays multiple rounds of games with winning configurations to ensure the performance guarantees of the selected tuning configuration. This lowers the performance variations (coefficient of variation of execution time) when \sol{} is integrated with ActiveHarmony and BLISS by 6.8\% and 5.9\%, respectively. 

\vspace{3mm}

\begin{figure}
    \centering
    \includegraphics[scale=0.31]{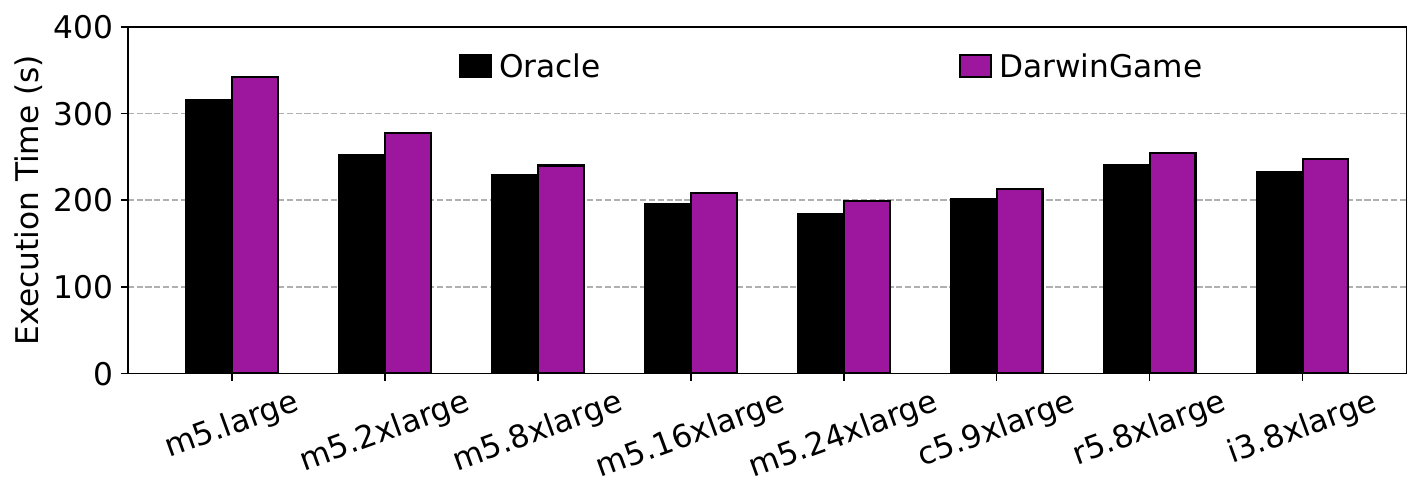}
    \vspace{-4mm}
    \caption{\sol{}'s effectiveness is consistent across different classes and sizes of VMs in the cloud (Redis serving one million requests).}
    \label{fig:tuner_diff_sys}
    \vspace{-2mm}
\end{figure}

\noindent\textbf{Are \sol{}'s benefits sensitive to or tied to one particular cloud-based VM instance type?} We perform experiments to determine \sol{}'s performance when both application tuning and execution are performed in different types and sizes of VMs. Different VMs have different amounts of interference (smaller VMs have higher interference due to the co-location of more tenants). 

In Fig.~\ref{fig:tuner_diff_sys} we observe that \sol{}'s execution time-wise performance always remains within 10\% of the Oracle solution. Also, the coefficient of variation of execution time with \sol{}'s chosen tuning configuration remains less than 0.46\%. These results experimentally confirm that \sol{}'s design components are generally applicable and the benefits are not tied to a particular VM type. 

\vspace{3mm}

\begin{figure}
    \centering
    \includegraphics[scale=0.33]{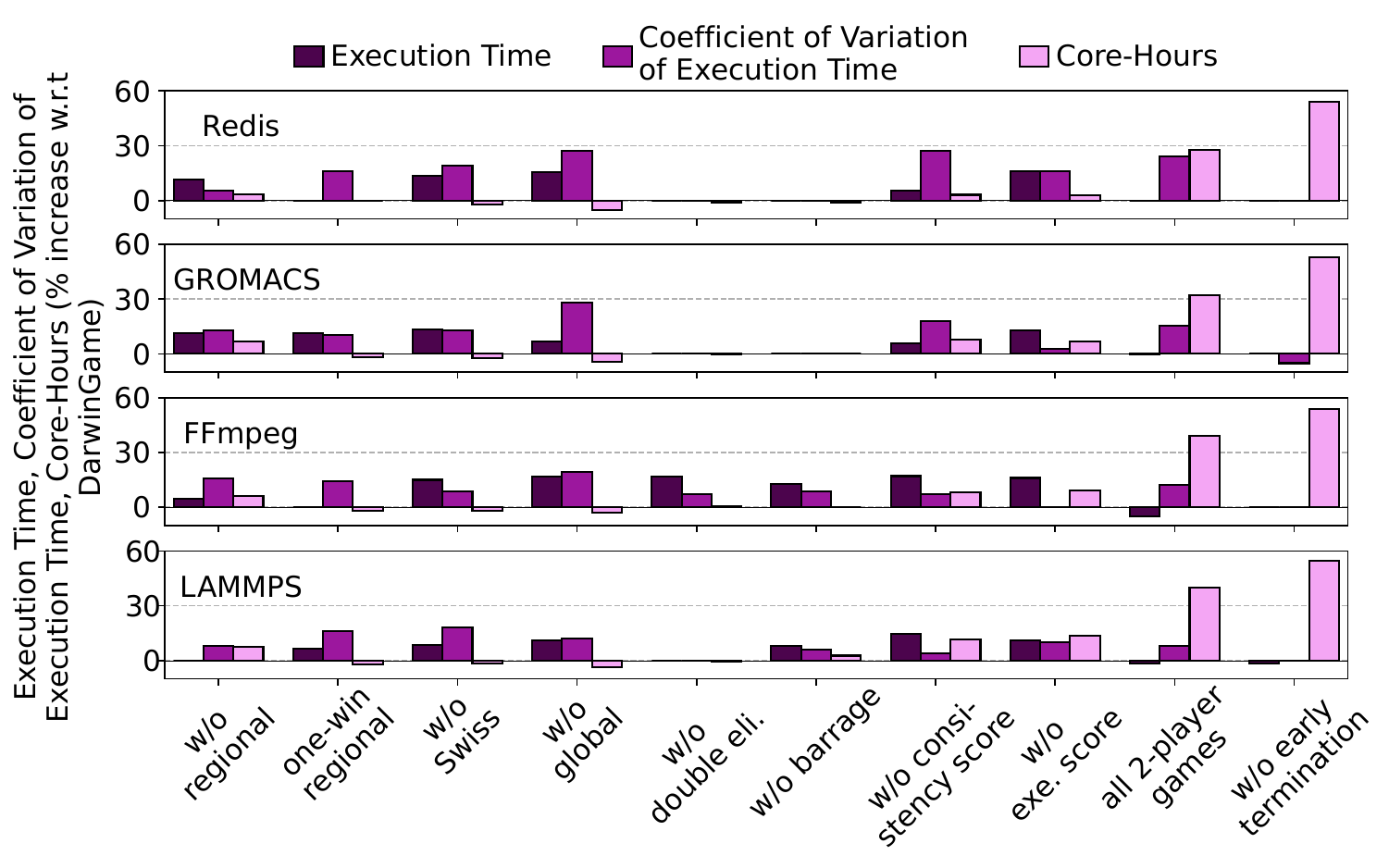}
    \vspace{-4mm}
    \caption{\sol{} performs better than alternative tournament structures.}
    \label{fig:tuner_alternate_all}
    \vspace{-2mm}
\end{figure}

\noindent\textbf{How does \sol{}'s tournament structure design help to achieve its effectiveness?} \sol{}'s tournament design consists of multiple tournament phases and gaming styles -- the first of them is the regional phase, which is played in Swiss style. In Fig.~\ref{fig:tuner_alternate_all}, we show the results for the case when \sol{} was played without a regional phase (w/o regional phase), and treating all players from the start in the global phase. We observe that the resultant performance in terms of execution time and performance variations deteriorate due to promising configurations getting eliminated early or not getting a chance to play. If in the regional phase, only one winner was selected per region (one-win regional), instead of \sol{}'s design of choosing the number of winners based on how good the overall regional performance is, then also the performance deteriorates due to promising players getting eliminated. Performance deterioration is also observed when the regional phase is not played in Swiss style (w/o Swiss), where only one game among all players selects the regional phase winners.

The next key design element in \sol{} is the global phase, played in double elimination style. Without the global phase (w/o global), \sol{}'s execution time and performance variations increase. In such a case, only one game is played among all regional winners to choose the players for the playoffs. We observe that for FFmpeg, without the double elimination (w/o double eli.) format, the execution time and performance variations increase, as the winner from the loser bracket finally wins the tournament for FFmpeg.  Without barrage-styled playoffs (w/o barrage), performance deteriorates for FFmpeg and LAMMPS as the loser of the first playoff actually wins the tournament. 

A notable key element in \sol{}'s design is the joint use of consistency and execution score. Using one of them, and not both (w/o consistency score, w/o exe. score), degrades \sol{}'s effectiveness, as both of these scores jointly determine the configuration that yields low execution time and consistent performance in the cloud. Without \sol{}'s design feature to reduce tuning time by playing games among multiple players in the initial phases (playing only two-player games -- all 2-player games), and without early termination (w/o early termination), \sol{}'s resource usage in terms of core-hours for tuning can increase by more than 30\%. Fig.~\ref{fig:tuner_alternate_all} summarizes how \sol{}'s design elements contribute to its effectiveness. 

\section{Related Works}
\label{sec:rel}

The complexity and importance of performance tuning in software systems are well-recognized~\cite{falch2015machine,
ogilvie2017minimizing,wu2019paraopt,ansel2012siblingrivalry,chatterjee2016declarative,chen2002tuning,li2024fasttuning}. Recent research has focused on automated methods: heuristic-based optimization like genetic algorithms and simulated annealing, machine learning approaches such as reinforcement learning and Bayesian optimization, and adaptive tuning systems~\cite{zhang2016maximizing,gschwandtner2014multi,kunen2015kripke,gerndt2010automatic, abdelfattah2016performance,rubio2013precimonious,guo2020pliner,silvano2016autotuning,chen2015angel,dotsenko2011auto,baghsorkhi2009analytical}. Heuristic methods offer structured ways to navigate search spaces but can be hindered by scalability and computational overheads, limiting effectiveness in dynamic environments like the cloud~\cite{ghasemi2024geyser,
huang2019survey}. Machine learning-based approaches predict optimal configurations efficiently, with Bayesian optimization suitable for high-dimensional settings despite needing extensive training data~\cite{roy2021bliss, liu2021gptune,souza2020prior,menon2020auto,dalibard2017boat,chen2018tvm,jamshidi2017transfer,grebhahn2016performance,
marathe2017performance}. Hybrid approaches, combining heuristics and machine learning, aim to balance exploration and exploitation but face challenges in computational cost and robustness~\cite{yu2020hyper}. Adaptive tuning systems continuously adjust configurations in real time using feedback loops and control theory. These systems are hindered by noise, unpredictability, and latency in responding to performance fluctuations~\cite{xu2024holistic,li2024accelerated,rasch2021efficient,balaprakash2018autotuning,tiwari2011online,behzad2019optimizing,raman2012parcae,cho2018maximizing}.

All existing tuning solutions rely on executing an application multiple times to determine the relative performance of an application execution with different tuning configurations. These solutions work when tuning is performed in a dedicated environment~\cite{roy2021bliss,ansel2014opentuner,wu2019paraopt,sourouri2017towards,yang2023cotuner,liu2021gptune,li2024accelerated}. However, when performing tuning in the cloud, a major obstacle is the inherent noise and variability in cloud environments, which can obscure the true performance impact of different configurations~\cite{de2022noise,ali2021hidden,zhu2019classytune}. Existing tuning methods assume a stable tuning environment, leading to suboptimal performance when used in production cloud settings. Moreover, there is a need for tuning solutions that can identify configurations with both low execution times and minimal performance variability. This gap underscores the necessity for novel approaches that can effectively account for and mitigate the impact of cloud-induced noise, ensuring reliable and consistent performance tuning.

\vspace{-4mm}
\section{Conclusion}
\sol{} is a novel tuning framework that effectively handles the noise and variability in cloud environments. \sol{} plays a tournament among multiple tuning configurations, with different phases of the tournaments customized to ensure finding tuning configurations with both low execution times and reduced performance variability. \sol{}'s evaluation demonstrates significant improvements -- achieving more than 27\% reduction in execution time compared to other existing solutions while limiting the performance variation to less than 0.5\%, marking a substantial advancement in the field.

\vspace{0mm}

\noindent\textbf{Acknowledgments.} We thank Cliff Young (our shepherd) and the other anonymous reviewers for their constructive feedback. The work is supported by the Assistant Secretary of Defense for Research and Engineering under Air Force
Contract No. FA8702-15-D-0001, and United States Air Force Research Laboratory Cooperative Agreement Number FA8750-19-2-
1000, and partially also supported by  by NSF Awards 1910601 and
2124897. Any opinions, findings, conclusions, or recommendations
expressed in this material are those of the author(s) and do not
necessarily reflect the views of the Assistant Secretary of Defense
for Research and Engineering, or the United States Air Force. 
The U.S. Government is authorized to reproduce and distribute reprints
for Government purposes notwithstanding any copyright notation
herein.




\balance
\bibliographystyle{plain}
\bibliography{main}

\begin{thebibliography}{10}

\bibitem{ucbnodesharing}
{ Berkeley Research Computing (BRC) Node Sharing}, url: {\url{https://docs-research-it.berkeley.edu/services/high-performance-computing/condos/condo-cluster-service/}}.

\bibitem{camberwellpetanqueclubMatchFormats}
{M}atch {F}ormats | {C}amberwell {P}etanque {C}lub --- camberwellpetanqueclub.org.au.
\newblock \url{https://www.camberwellpetanqueclub.org.au/content/match-formats}.

\bibitem{utahhpc}
{Node Sharing Policy at CHPC, Utah}, url: {\url{https://www.chpc.utah.edu/documentation/software/node-sharing.php}}.

\bibitem{sdsccondo}
{SDSC Condo HPC model}, url: {\url{https://www.hpcwire.com/2020/08/10/high-performance-computing-at-condo-prices/}}.

\bibitem{key}
Tournament formats -- an overview.
\newblock \url{https://petanquerules.wordpress.com/tournament-systems}.

\bibitem{abdelfattah2016performance}
Ahmad Abdelfattah, Azzam Haidar, Stanimire Tomov, and Jack Dongarra.
\newblock {Performance, Design, and Autotuning of Batched GEMM for GPUs}.
\newblock In {\em International Conference on High Performance Computing}, pages 21--38. Springer, 2016.

\bibitem{abraham2015gromacs}
Mark~James Abraham, Teemu Murtola, Roland Schulz, Szil{\'a}rd P{\'a}ll, Jeremy~C Smith, Berk Hess, and Erik Lindahl.
\newblock Gromacs: High performance molecular simulations through multi-level parallelism from laptops to supercomputers.
\newblock {\em SoftwareX}, 1:19--25, 2015.

\bibitem{akiba2019optuna}
Takuya Akiba, Shotaro Sano, Toshihiko Yanase, Takeru Ohta, and Masanori Koyama.
\newblock Optuna: A next-generation hyperparameter optimization framework.
\newblock In {\em Proceedings of the 25th ACM SIGKDD international conference on knowledge discovery \& data mining}, pages 2623--2631, 2019.

\bibitem{alam2020cloud}
Tanweer Alam.
\newblock Cloud computing and its role in the information technology.
\newblock {\em IAIC Transactions on Sustainable Digital Innovation (ITSDI)}, 1(2):108--115, 2020.

\bibitem{ali2021hidden}
Ahmed Ali-Eldin, Bin Wang, and Prashant Shenoy.
\newblock The hidden cost of the edge: a performance comparison of edge and cloud latencies.
\newblock In {\em Proceedings of the International Conference for High Performance Computing, Networking, Storage and Analysis}, pages 1--12, 2021.

\bibitem{ansel2014opentuner}
Jason Ansel, Shoaib Kamil, Kalyan Veeramachaneni, Jonathan Ragan-Kelley, Jeffrey Bosboom, Una-May O'Reilly, and Saman Amarasinghe.
\newblock Opentuner: An extensible framework for program autotuning.
\newblock In {\em Proceedings of the 23rd international conference on Parallel architectures and compilation}, pages 303--316, 2014.

\bibitem{ansel2012siblingrivalry}
Jason Ansel, Maciej Pacula, Yee~Lok Wong, Cy~Chan, Marek Olszewski, Una-May O'Reilly, and Saman Amarasinghe.
\newblock Siblingrivalry: online autotuning through local competitions.
\newblock In {\em Proceedings of the 2012 international conference on Compilers, architectures and synthesis for embedded systems}, pages 91--100, 2012.

\bibitem{baghsorkhi2009analytical}
Sara~S Baghsorkhi, Matthieu Delahaye, William~D Gropp, and W~Hwu Wen-mei.
\newblock Analytical performance prediction for evaluation and tuning of gpgpu applications.
\newblock In {\em Workshop on EPHAM2009}, 2009.

\bibitem{balaprakash2018autotuning}
Prasanna Balaprakash, Jack Dongarra, Todd Gamblin, Mary Hall, Jeffrey~K Hollingsworth, Boyana Norris, and Richard Vuduc.
\newblock Autotuning in high-performance computing applications.
\newblock {\em Proceedings of the IEEE}, 106(11):2068--2083, 2018.

\bibitem{bao2019actgan}
Liang Bao, Xin Liu, Fangzheng Wang, and Baoyin Fang.
\newblock Actgan: automatic configuration tuning for software systems with generative adversarial networks.
\newblock In {\em 2019 34th IEEE/ACM International Conference on Automated Software Engineering (ASE)}, pages 465--476. IEEE, 2019.

\bibitem{behzad2019optimizing}
Babak Behzad, Surendra Byna, Prabhat, and Marc Snir.
\newblock Optimizing i/o performance of hpc applications with autotuning.
\newblock {\em ACM Transactions on Parallel Computing (TOPC)}, 5(4):1--27, 2019.

\bibitem{bronevetsky2008clomp}
Greg Bronevetsky, John Gyllenhaal, and Bronis~R De~Supinski.
\newblock {CLOMP: Accurately Characterizing OpenMP Application Overheads}.
\newblock In {\em International Workshop on OpenMP}, pages 13--25. Springer, 2008.

\bibitem{buyya2018manifesto}
Rajkumar Buyya, Satish~Narayana Srirama, Giuliano Casale, Rodrigo Calheiros, Yogesh Simmhan, Blesson Varghese, Erol Gelenbe, Bahman Javadi, Luis~Miguel Vaquero, Marco~AS Netto, et~al.
\newblock A manifesto for future generation cloud computing: Research directions for the next decade.
\newblock {\em ACM computing surveys (CSUR)}, 51(5):1--38, 2018.

\bibitem{carlson2013redis}
Josiah Carlson.
\newblock {\em Redis in action}.
\newblock Simon and Schuster, 2013.

\bibitem{chatterjee2016declarative}
Sanjay Chatterjee, Nick Vrvilo, Zoran Budimlic, Kathleen Knobe, and Vivek Sarkar.
\newblock Declarative tuning for locality in parallel programs.
\newblock In {\em 45th International Conference on Parallel Processing (ICPP)}, 2016.

\bibitem{chen2002tuning}
Guangyu Chen et~al.
\newblock Tuning garbage collection in an embedded java environment.
\newblock In {\em HPCA}. IEEE, 2002.

\bibitem{chen2015angel}
Ray~S Chen and Jeffrey~K Hollingsworth.
\newblock {Angel: A Hierarchical Approach to Multi-Objective Online Auto-Tuning}.
\newblock In {\em Proceedings of the 5th International Workshop on Runtime and Operating Systems for Supercomputers}, pages 1--8, 2015.

\bibitem{chen2018tvm}
Tianqi Chen, Thierry Moreau, Ziheng Jiang, Lianmin Zheng, Eddie Yan, Haichen Shen, Meghan Cowan, Leyuan Wang, Yuwei Hu, Luis Ceze, et~al.
\newblock {$\{$TVM$\}$: An Automated End-to-End Optimizing Compiler for Deep Learning}.
\newblock In {\em 13th $\{$USENIX$\}$ Symposium on Operating Systems Design and Implementation ($\{$OSDI$\}$ 18)}, pages 578--594, 2018.

\bibitem{cho2018maximizing}
Younghyun Cho, Camilo A~Celis Guzman, and Bernhard Egger.
\newblock Maximizing system utilization via parallelism management for co-located parallel applications.
\newblock In {\em Proceedings of the 27th International Conference on Parallel Architectures and Compilation Techniques}, pages 1--14, 2018.

\bibitem{copik2024software}
Marcin Copik, Marcin Chrapek, Larissa Schmid, Alexandru Calotoiu, and Torsten Hoefler.
\newblock Software resource disaggregation for hpc with serverless computing.
\newblock {\em arXiv preprint arXiv:2401.10852}, 2024.

\bibitem{csato2021tournament}
L{\'a}szl{\'o} Csat{\'o}.
\newblock {\em Tournament design: How operations research can improve sports rules}.
\newblock Springer Nature, 2021.

\bibitem{dalibard2017boat}
Valentin Dalibard, Michael Schaarschmidt, and Eiko Yoneki.
\newblock Boat: Building auto-tuners with structured bayesian optimization.
\newblock In {\em WWW}, pages 479--488, 2017.

\bibitem{de2022noise}
Daniele De~Sensi, Tiziano De~Matteis, Konstantin Taranov, Salvatore Di~Girolamo, Tobias Rahn, and Torsten Hoefler.
\newblock Noise in the clouds: Influence of network performance variability on application scalability.
\newblock {\em Proceedings of the ACM on Measurement and Analysis of Computing Systems}, 6(3):1--27, 2022.

\bibitem{delimitrou2016hcloud}
Christina Delimitrou and Christos Kozyrakis.
\newblock Hcloud: Resource-efficient provisioning in shared cloud systems.
\newblock In {\em Proceedings of the Twenty-First International Conference on Architectural Support for Programming Languages and Operating Systems}, pages 473--488, 2016.

\bibitem{destefano2023cloud}
Timothy DeStefano, Richard Kneller, and Jonathan Timmis.
\newblock Cloud computing and firm growth.
\newblock {\em Review of Economics and Statistics}, pages 1--47, 2023.

\bibitem{ding2015autotuning}
Yufei Ding et~al.
\newblock Autotuning algorithmic choice for input sensitivity.
\newblock In {\em ACM SIGPLAN Conference on Programming Language Design and Implementation (PLDI)}, 2015.

\bibitem{dinh2020simulating}
An~Vinh~Nguyen Dinh, Nhien Pham~Hoang Bao, Mohd Nor~Akmal Khalid, and Hiroyuki Iida.
\newblock Simulating competitiveness and precision in a tournament structure: A reaper tournament system.
\newblock {\em International Journal of Information Technology}, 12:1--18, 2020.

\bibitem{dotsenko2011auto}
Yuri Dotsenko, Sara~S Baghsorkhi, Brandon Lloyd, and Naga~K Govindaraju.
\newblock Auto-tuning of fast fourier transform on graphics processors.
\newblock {\em ACM SIGPLAN Notices}, 46(8):257--266, 2011.

\bibitem{falch2015machine}
Thomas~L Falch and Anne~C Elster.
\newblock {Machine Learning based Auto-Tuning for Enhanced OpenCL Performance Portability}.
\newblock In {\em 2015 IEEE International Parallel and Distributed Processing Symposium Workshop}, pages 1231--1240. IEEE, 2015.

\bibitem{fuhrlich2022improving}
Pascal F{\"u}hrlich, {\'A}gnes Cseh, and Pascal Lenzner.
\newblock Improving ranking quality and fairness in swiss-system chess tournaments.
\newblock In {\em Proceedings of the 23rd ACM Conference on Economics and Computation}, pages 1101--1102, 2022.

\bibitem{gerndt2010automatic}
Michael Gerndt and Michael Ott.
\newblock {Automatic Performance Analysis with Periscope}.
\newblock {\em Concurrency and Computation: Practice and Experience}, 22(6):736--748, 2010.

\bibitem{ghasemi2024geyser}
Mojtaba Ghasemi, Mohsen Zare, Amir Zahedi, Mohammad-Amin Akbari, Seyedali Mirjalili, and Laith Abualigah.
\newblock Geyser inspired algorithm: a new geological-inspired meta-heuristic for real-parameter and constrained engineering optimization.
\newblock {\em Journal of Bionic Engineering}, 21(1):374--408, 2024.

\bibitem{gratl2019autopas}
Fabio~Alexander Gratl, Steffen Seckler, Nikola Tchipev, Hans-Joachim Bungartz, and Philipp Neumann.
\newblock Autopas: Auto-tuning for particle simulations.
\newblock In {\em 2019 IEEE International Parallel and Distributed Processing Symposium Workshops (IPDPSW)}, pages 748--757. IEEE, 2019.

\bibitem{grebhahn2016performance}
Alexander Grebhahn, Norbert Siegmund, Harald K{\"o}stler, and Sven Apel.
\newblock {Performance Prediction of Multigrid-Solver Configurations}.
\newblock In {\em Software for Exascale Computing-SPPEXA 2013-2015}. Springer, 2016.

\bibitem{gschwandtner2014multi}
Philipp Gschwandtner et~al.
\newblock {Multi-Objective Auto-Tuning with Insieme: Optimization and Trade-Off Analysis for Time, Energy and Resource Usage}.
\newblock In {\em European Conference on Parallel Processing}, 2014.

\bibitem{guo2020pliner}
Hui Guo, Ignacio Laguna, and Cindy Rubio-Gonz{\'a}lez.
\newblock pliner: isolating lines of floating-point code for compiler-induced variability.
\newblock In {\em International Conference for High Performance Computing, Networking, Storage and Analysis (SC)}, pages 680--693. IEEE Computer Society, 2020.

\bibitem{gupta2011evaluation}
Abhishek Gupta and Dejan Milojicic.
\newblock Evaluation of hpc applications on cloud.
\newblock In {\em 2011 Sixth Open Cirrus Summit}, pages 22--26. IEEE, 2011.

\bibitem{hampton2010optimal}
Scott~S Hampton, Sadaf~R Alam, Paul~S Crozier, and Pratul~K Agarwal.
\newblock Optimal utilization of heterogeneous resources for biomolecular simulations.
\newblock In {\em SC'10: Proceedings of the 2010 ACM/IEEE International Conference for High Performance Computing, Networking, Storage and Analysis}, pages 1--11. IEEE, 2010.

\bibitem{harbring2003experimental}
Christine Harbring and Bernd Irlenbusch.
\newblock An experimental study on tournament design.
\newblock {\em Labour Economics}, 10(4):443--464, 2003.

\bibitem{hollingsworth2010end}
Jeffrey Hollingsworth and Ananta Tiwari.
\newblock End-to-end auto-tuning with active harmony.
\newblock {\em Performance tuning of scientific applications}, pages 217--238, 2010.

\bibitem{hollingsworth1999prediction}
Jeffrey~K Hollingsworth and Peter~J Keleher.
\newblock {Prediction and Adaptation in Active Harmony}.
\newblock {\em Cluster Computing}, 2(3):195, 1999.

\bibitem{hua2017swiss}
Christopher Hua.
\newblock The swiss tournament model.
\newblock 2017.

\bibitem{huang2019survey}
Changwu Huang, Yuanxiang Li, and Xin Yao.
\newblock A survey of automatic parameter tuning methods for metaheuristics.
\newblock {\em IEEE transactions on evolutionary computation}, 24(2):201--216, 2019.

\bibitem{jamshidi2017transfer}
Pooyan Jamshidi, Miguel Velez, Christian K{\"a}stner, Norbert Siegmund, and Prasad Kawthekar.
\newblock {Transfer Learning for Improving Model Predictions in Highly Configurable Software}.
\newblock In {\em 2017 IEEE/ACM 12th International Symposium on Software Engineering for Adaptive and Self-Managing Systems}, 2017.

\bibitem{klein2014brownout}
Cristian Klein, Martina Maggio, Karl-Erik {\AA}rz{\'e}n, and Francisco Hern{\'a}ndez-Rodriguez.
\newblock Brownout: Building more robust cloud applications.
\newblock In {\em Proceedings of the 36th International Conference on Software Engineering}, pages 700--711, 2014.

\bibitem{kunen2015kripke}
Adam~J Kunen, Teresa~S Bailey, and Peter~N Brown.
\newblock {KRIPKE: A Massively Parallel Transport Mini-App}.
\newblock Technical report, Lawrence Livermore National Lab.(LLNL), Livermore, CA (United States), 2015.

\bibitem{li2024accelerated}
Chendi Li, Yufan Xu, Sina~Mahdipour Saravani, and Ponnuswamy Sadayappan.
\newblock Accelerated auto-tuning of gpu kernels for tensor computations.
\newblock In {\em Proceedings of the 38th ACM International Conference on Supercomputing}, pages 549--561, 2024.

\bibitem{li2024fasttuning}
Xiaqing Li, Qi~Guo, Guangyan Zhang, Siwei Ye, Guanhua He, Yiheng Yao, Rui Zhang, Yifan Hao, Zidong Du, and Weimin Zheng.
\newblock Fasttuning: Enabling fast and efficient hyper-parameter tuning with partitioning and parallelism of search space.
\newblock {\em IEEE Transactions on Parallel and Distributed Systems}, 2024.

\bibitem{liu2021gptune}
Yang Liu, Wissam~M Sid-Lakhdar, Osni Marques, Xinran Zhu, Chang Meng, James~W Demmel, and Xiaoye~S Li.
\newblock Gptune: multitask learning for autotuning exascale applications.
\newblock In {\em Proceedings of the 26th ACM SIGPLAN Symposium on Principles and Practice of Parallel Programming}, pages 234--246, 2021.

\bibitem{marathe2017performance}
Aniruddha Marathe et~al.
\newblock {Performance Modeling Under Resource Constraints Using Deep Transfer Learning}.
\newblock In {\em Proceedings of the International Conference for High Performance Computing, Networking, Storage and Analysis}, pages 1--12, 2017.

\bibitem{menon2020auto}
Harshitha Menon, Abhinav Bhatele, and Todd Gamblin.
\newblock Auto-tuning parameter choices in hpc applications using bayesian optimization.
\newblock In {\em IPDPS}, pages 831--840. IEEE, 2020.

\bibitem{ogilvie2017minimizing}
William~F Ogilvie, Pavlos Petoumenos, Zheng Wang, and Hugh Leather.
\newblock {Minimizing the Cost of Iterative Compilation with Active Learning}.
\newblock In {\em 2017 IEEE/ACM International Symposium on Code Generation and Optimization (CGO)}, 2017.

\bibitem{rajkumar2021theory}
Arun Rajkumar, Vishnu Veerathu, and Abdul~Bakey Mir.
\newblock A theory of tournament representations.
\newblock {\em arXiv preprint arXiv:2110.05188}, 2021.

\bibitem{raman2012parcae}
Arun Raman, Ayal Zaks, Jae~W Lee, and David~I August.
\newblock Parcae: a system for flexible parallel execution.
\newblock {\em ACM SIGPLAN Notices}, 47(6):133--144, 2012.

\bibitem{rasch2021efficient}
Ari Rasch, Richard Schulze, Michel Steuwer, and Sergei Gorlatch.
\newblock Efficient auto-tuning of parallel programs with interdependent tuning parameters via auto-tuning framework (atf).
\newblock {\em ACM Transactions on Architecture and Code Optimization (TACO)}, 18(1):1--26, 2021.

\bibitem{reuther2020survey}
Albert Reuther, Peter Michaleas, Michael Jones, Vijay Gadepally, Siddharth Samsi, and Jeremy Kepner.
\newblock Survey of machine learning accelerators.
\newblock In {\em 2020 IEEE high performance extreme computing conference (HPEC)}, pages 1--12. IEEE, 2020.

\bibitem{roy2021bliss}
Rohan~Basu Roy, Tirthak Patel, Vijay Gadepally, and Devesh Tiwari.
\newblock Bliss: auto-tuning complex applications using a pool of diverse lightweight learning models.
\newblock In {\em Proceedings of the 42nd ACM SIGPLAN International Conference on Programming Language Design and Implementation}, pages 1280--1295, 2021.

\bibitem{rubio2013precimonious}
Cindy Rubio-Gonz{\'a}lez, Cuong Nguyen, Hong~Diep Nguyen, James Demmel, William Kahan, Koushik Sen, David~H Bailey, Costin Iancu, and David Hough.
\newblock Precimonious: Tuning assistant for floating-point precision.
\newblock In {\em SC}, 2013.

\bibitem{ryvkin2008predictive}
Dmitry Ryvkin and Andreas Ortmann.
\newblock The predictive power of three prominent tournament formats.
\newblock {\em Management Science}, 54(3):492--504, 2008.

\bibitem{sarkar2020ddaring}
Vivek Sarkar.
\newblock Ddaring: Dynamic data aware reconfiguration, integration and generation.
\newblock Technical report, GEORGIA TECH RESEARCH CORPORATION Atlanta United States, 2020.

\bibitem{sarkar2009software}
Vivek Sarkar, William Harrod, and Allan~E Snavely.
\newblock Software challenges in extreme scale systems.
\newblock In {\em Journal of Physics: Conference Series}. IOP Publishing, 2009.

\bibitem{shafiei2022serverless}
Hossein Shafiei, Ahmad Khonsari, and Payam Mousavi.
\newblock Serverless computing: a survey of opportunities, challenges, and applications.
\newblock {\em ACM Computing Surveys}, 54(11s):1--32, 2022.

\bibitem{silvano2016antarex}
Cristina Silvano, Giovanni Agosta, Stefano Cherubin, Davide Gadioli, Gianluca Palermo, Andrea Bartolini, Luca Benini, Jan Martinovi{\v{c}}, Martin Palkovi{\v{c}}, Kate{\v{r}}ina Slaninov{\'a}, et~al.
\newblock The antarex approach to autotuning and adaptivity for energy efficient hpc systems.
\newblock In {\em Proceedings of the ACM International Conference on Computing Frontiers}, pages 288--293, 2016.

\bibitem{silvano2016autotuning}
Cristina Silvano et~al.
\newblock {AutoTuning and Adaptivity AppRoach for Energy Efficient EXascale HPC Systems: The ANTAREX Approach}.
\newblock In {\em 2016 Design, Automation \& Test in Europe Conference \& Exhibition (DATE)}, pages 708--713. IEEE, 2016.

\bibitem{simakov2016quantitative}
Nikolay~A Simakov, Robert~L DeLeon, Joseph~P White, Thomas~R Furlani, Martins Innus, Steven~M Gallo, Matthew~D Jones, Abani Patra, Benjamin~D Plessinger, Jeanette Sperhac, et~al.
\newblock A quantitative analysis of node sharing on hpc clusters using xdmod application kernels.
\newblock In {\em Proceedings of the XSEDE16 Conference on Diversity, Big Data, and Science at Scale}, pages 1--8, 2016.

\bibitem{sourouri2017towards}
Mohammed Sourouri, Espen~Birger Raknes, Nico Reissmann, Johannes Langguth, Daniel Hackenberg, Robert Sch{\"o}ne, and Per~Gunnar Kjeldsberg.
\newblock Towards fine-grained dynamic tuning of hpc applications on modern multi-core architectures.
\newblock In {\em Proceedings of the International Conference for High Performance Computing, Networking, Storage and Analysis}, pages 1--12, 2017.

\bibitem{souza2020prior}
Artur Souza, Luigi Nardi, Leonardo~B Oliveira, Kunle Olukotun, Marius Lindauer, and Frank Hutter.
\newblock Prior-guided bayesian optimization.
\newblock {\em arXiv preprint arXiv:2006.14608}, 2020.

\bibitem{tapus2002active}
Cristian Tapus, I-Hsin Chung, and Jeffrey~K Hollingsworth.
\newblock {Active Harmony: Towards Automated Performance Tuning}.
\newblock In {\em Proceedings of the 2002 ACM/IEEE Conference on Supercomputing (SC)}, 2002.

\bibitem{thiagarajan2018bootstrapping}
Jayaraman~J Thiagarajan, Nikhil Jain, Rushil Anirudh, Alfredo Gimenez, Rahul Sridhar, Aniruddha Marathe, Tao Wang, Murali Emani, Abhinav Bhatele, and Todd Gamblin.
\newblock Bootstrapping parameter space exploration for fast tuning.
\newblock In {\em Proceedings of the 2018 International Conference on Supercomputing}, pages 385--395, 2018.

\bibitem{thompson2022lammps}
Aidan~P Thompson, H~Metin Aktulga, Richard Berger, Dan~S Bolintineanu, W~Michael Brown, Paul~S Crozier, Pieter~J In't~Veld, Axel Kohlmeyer, Stan~G Moore, Trung~Dac Nguyen, et~al.
\newblock Lammps-a flexible simulation tool for particle-based materials modeling at the atomic, meso, and continuum scales.
\newblock {\em Computer Physics Communications}, 271:108171, 2022.

\bibitem{tiwari2011auto}
Ananta Tiwari et~al.
\newblock Auto-tuning full applications: A case study.
\newblock {\em The International Journal of High Performance Computing Applications}, 25(3):286--294, 2011.

\bibitem{tiwari2011online}
Ananta Tiwari et~al.
\newblock {Online Adaptive Code Generation and Tuning}.
\newblock In {\em 2011 IEEE International Parallel \& Distributed Processing Symposium}, pages 879--892. IEEE, 2011.

\bibitem{tomar2006converting}
Suramya Tomar.
\newblock Converting video formats with ffmpeg.
\newblock {\em Linux journal}, 2006(146):10, 2006.

\bibitem{varghese2018next}
Blesson Varghese and Rajkumar Buyya.
\newblock Next generation cloud computing: New trends and research directions.
\newblock {\em Future Generation Computer Systems}, 79:849--861, 2018.

\bibitem{wittig2023amazon}
Andreas Wittig and Michael Wittig.
\newblock {\em Amazon Web Services in Action: An in-depth guide to AWS}.
\newblock Simon and Schuster, 2023.

\bibitem{wu2019paraopt}
Chaofeng Wu et~al.
\newblock Paraopt: Automated application parameterization and optimization for the cloud.
\newblock In {\em 2019 IEEE International Conference on Cloud Computing Technology and Science (CloudCom)}, pages 255--262. IEEE, 2019.

\bibitem{wu2023ytopt}
Xingfu Wu, Prasanna Balaprakash, Michael Kruse, Jaehoon Koo, Brice Videau, Paul Hovland, Valerie Taylor, Brad Geltz, Siddhartha Jana, and Mary Hall.
\newblock ytopt: Autotuning scientific applications for energy efficiency at large scales.
\newblock {\em arXiv preprint arXiv:2303.16245}, 2023.

\bibitem{xiong2018tangram}
Qingqing Xiong, Emre Ates, Martin~C Herbordt, and Ayse~K Coskun.
\newblock Tangram: Colocating hpc applications with oversubscription.
\newblock In {\em 2018 IEEE High Performance extreme Computing Conference (HPEC)}, pages 1--7. IEEE, 2018.

\bibitem{xu2024holistic}
Jinchen Xu, Guanghui Song, Bei Zhou, Fei Li, Jiangwei Hao, and Jie Zhao.
\newblock A holistic approach to automatic mixed-precision code generation and tuning for affine programs.
\newblock In {\em Proceedings of the 29th ACM SIGPLAN Annual Symposium on Principles and Practice of Parallel Programming}, pages 55--67, 2024.

\bibitem{yang2023cotuner}
Tiannuo Yang, Ruobing Chen, Yusen Li, Xiaoguang Liu, and Gang Wang.
\newblock Cotuner: A hierarchical learning framework for coordinately optimizing resource partitioning and parameter tuning.
\newblock In {\em Proceedings of the 52nd International Conference on Parallel Processing}, pages 317--326, 2023.

\bibitem{yu2020hyper}
Tong Yu and Hong Zhu.
\newblock Hyper-parameter optimization: A review of algorithms and applications.
\newblock {\em arXiv preprint arXiv:2003.05689}, 2020.

\bibitem{zhang2016maximizing}
Huazhe Zhang and Henry Hoffmann.
\newblock {Maximizing Performance under a Power Cap: A Comparison of Hardware, Software, and Hybrid Techniques}.
\newblock {\em ACM SIGPLAN Notices}, 51(4):545--559, 2016.

\bibitem{zhang2020high}
Xiantao Zhang, Xiao Zheng, Zhi Wang, Hang Yang, Yibin Shen, and Xin Long.
\newblock High-density multi-tenant bare-metal cloud.
\newblock In {\em Proceedings of the Twenty-Fifth International Conference on Architectural Support for Programming Languages and Operating Systems}, pages 483--495, 2020.

\bibitem{zhu2019classytune}
Yuqing Zhu and Jianxun Liu.
\newblock Classytune: A performance auto-tuner for systems in the cloud.
\newblock {\em IEEE Transactions on Cloud Computing}, 10(1):234--246, 2019.

\end{thebibliography}

\end{document}